\documentclass[sigconf]{acmart}
\usepackage{multirow}
\usepackage{enumitem}
\usepackage{balance}
\usepackage{graphicx}
\usepackage{subcaption}
\usepackage{cleveref}   
\AtBeginDocument{%
  \providecommand\BibTeX{{%
    \normalfont B\kern-0.5em{\scshape i\kern-0.25em b}\kern-0.8em\TeX}}}

\copyrightyear{2022}
\acmYear{2022}
\setcopyright{acmcopyright}
\acmConference[KDD '22] {Proceedings of the 28th ACM SIGKDD Conference on Knowledge Discovery and Data Mining}{August 14--18, 2022}{Washington, DC, USA.}
\acmBooktitle{Proceedings of the 28th ACM SIGKDD Conference on Knowledge Discovery and Data Mining (KDD '22), August 14--18, 2022, Washington, DC, USA}
\acmPrice{15.00}
\acmDOI{10.1145/3534678.3539239}
\acmISBN{978-1-4503-9385-0/22/08}

\begin{document}

\title{MetroGAN: Simulating Urban Morphology with Generative Adversarial Network}

\author{Weiyu Zhang}
\orcid{1234-5678-9012}
\affiliation{%
  \institution{Institute of Remote Sensing and Geographic Information Systems, School of Earth and Space Sciences, Peking University}
  \city{Beijing}
  \country{China}
}
\email{wyzhang929@gmail.com}

\author{Yiyang Ma}
\affiliation{%
 \institution{Wangxuan Institute of Computer Technology, Peking University}
 \city{Beijing}
 \country{China}}
\email{myy12769@pku.edu.cn}
 
\author{Di Zhu}
\affiliation{%
  \institution{Department of Geography, Environment and Society, University of Minnesota, Twin Cities}
  \city{Minneapolis}
  \country{USA}}
\email{dizhu@umn.edu}

\author{Lei Dong}
\affiliation{%
  \institution{Institute of Remote Sensing and Geographic Information Systems, School of Earth and Space Sciences, Peking University}
  \city{Beijing}
  \country{China}
}
\email{arch.dongl@gmail.com}

\author{Yu Liu}
\authornote{Contact author.}
\affiliation{%
  \institution{Institude of Remote Sensing and Geographic Information Systems, School of Earth and Space Sciences, Peking University}
  \city{Beijing}
  \country{China}}
\email{liuyu@urban.pku.edu.cn}

\renewcommand{\shortauthors}{Weiyu Zhang et al.}
\begin{abstract}
Simulating urban morphology with location attributes is a challenging task in urban science. Recent studies have shown that Generative Adversarial Networks (GANs) have the potential to shed light on this task. However, existing GAN-based models are limited by the sparsity of urban data and instability in model training, hampering their applications. Here, we propose a GAN framework with geographical knowledge, namely Metropolitan GAN  (MetroGAN), for urban morphology simulation. We incorporate a progressive growing structure to learn hierarchical features and design a geographical loss to impose the constraints of water areas. Besides, we propose a comprehensive evaluation framework for the complex structure of urban systems. Results show that MetroGAN outperforms the state-of-the-art urban simulation methods by over 20\% in all metrics. Inspiringly, using physical geography features singly, MetroGAN can still generate shapes of the cities. These results demonstrate that MetroGAN solves the instability problem of previous urban simulation GANs and is generalizable to deal with various urban attributes.

\end{abstract}


\maketitle

\section{Introduction}
The dynamics of cities influences many fields of our society, including economic growth, climate change, and sustainable development \cite{GrowthInnovationScalingPaceLife2007-Bettencourt}. To study urban dynamics, morphology or the spatial configuration is the key dimension \cite{SizeScaleShapeCities2008-Batty,batty1994fractal}, which not only provides a delineative representation of cities, but also is closely related to the complex characteristics of cities, such as the fractality, the polycentricity, and the scaling laws \cite{batty1994fractal,liSimpleSpatialScaling2017}.

Among relevant studies on urban morphology, simulating the spatial structure of cities is one critical task. Existing work addresses this problem in two main ways. On the one hand, studies represented by statistical physics \cite{EmergenceUrbanGrowthPatternsHuman2021-Xu} or complexity science (such as cellular automata \cite{liuSimulatingUrbanGrowth2014}) usually model city growth from a bottom-up perspective \cite{EmergenceUrbanGrowthPatternsHuman2021-Xu}. These models can capture key growth mechanisms of cities, but the generated form is often very different from the real situation because of ignoring many complex high-dimensional features of urban morphology \cite{liuSimulatingUrbanGrowth2014}. On the other hand, machine learning methods, especially deep generative models, have been validated owning the ability to approximate complicated, high-dimensional probability distributions \cite{GenerativeAdversarialNets2014-Goodfellow} and have been widely used in urban science and geoscience in recent years  \cite{SpatialInterpolationUsingConditionalGenerative2020-Zhua,ravuriSkilfulPrecipitationNowcasting2021}. Among deep generative models, the Generative Adversarial Network (GAN) is proved to be a suitable method to study spatial effects of geographical systems \cite{ModelingUrbanizationPatternsGenerativeAdversarial2018-Alberta, SpatialSensitivityAnalysisUrbanLand2019-Albert}. Specifically, generating urban morphology is similar to the task of image generation, which has been widely investigated in the GAN model family. However, the main problem in the urban setting is the severely sparse data (pixels with zero or low values account for a large proportion), which makes models hard to train. The performances of previous GAN-based models are limited by such data characteristics and face instability problems. 

An important way to solve the problem of data shortage and improve the performance of the model is to incorporate geographical knowledge into models \cite{PhysicsinformedMachineLearning2021-Karniadakis}. Based on this idea, we design a GAN-based model for urban morphology simulation, namely the Metropolitan GAN (MetroGAN). Considering the hierarchical and self-similar characteristics of urban systems, we introduce progressive growing structure into our model. This structure can gradually learn the urban features at different scales and significantly stabilize the model. Besides, we design a geographical loss to approximate the constraint that urban areas can not be on the water, making our model easier to learn the impacts of water areas.

To generate cities with MetroGAN, we collected the urban built-up area map (one commonly used proxy of urban morphology) as labels in our experiments. For inputs, we build a global city dataset using three layers: terrain [digital elevation model (DEM)], water, and nighttime lights (NTL). The first two are location's physical geography characteristics, which might dominate the growth of cities \cite{CitySeedsGeographyOriginsEuropean2017-Bosker}. NTL is a good proxy of socioeconomic development and has been recognized to be helpful in city extracting tasks \cite{UrbanMappingUsingDMSPOLS2017-Li,QuantifyingUrbanAreasMultisourceData2020-Cao}. Based on these three layers, we detect over 10,000 cities worldwide, and each city is represented as a 100km $\times$ 100km image, covering the central urban districts and surrounding settlements. 

Since a city is a complex system with plenty of structural and hierarchical features, we also propose an evaluation framework to assess the generated cities. Conventional evaluations always use pixel-by-pixel accuracy, which omit the connections among pixels and therefore are not sufficient to measure urban morphology. Our evaluation framework consists of four levels: pixel level, multi-scale spatial level, perceptual level, and macroscopic level. The first three levels are proposed for comparison. The multi-scale spatial level is added to consider the spatial structure of the urban system, and the perceptual level aims to compare the underlying structured information of urban form. The macroscopic level is for validation using geographical characteristics and laws.   

Results show that using NTL, terrain, and water area as inputs, our model can generate precise urban morphology. The generated cities improve around 20\% performance in terms of four metrics listed in Section 5 compared with state-of-art models. Furthermore, with physical geography characteristics singly, our model can still generate rough shapes of cities. 

Our model has a wide range of applications in urban science. For example, by learning the mappings from attributes to morphology, our model could help researchers explore the relations between location attributes and urban morphology. Further, our model provides a powerful tool for urban data augmentation. We can reconstruct urban change over a longer time series with limited snapshots.

To summarize, this research mainly has four contributions:
\begin{itemize}[leftmargin=*]
    \item We introduce MetroGAN, a model to simulate cities. The model incorporates the urban morphology with progressive growing structures, and physical geography constraints with designed geographical loss. 
    \item Our model can generate cities' shapes just with physical geography characteristics, shedding light on how and to what extent nature environment constraints could affect the cities.
    \item We propose a comprehensive evaluation framework for assessing urban morphology, which can be used for general city extraction, prediction, and simulation tasks.
    \item We publish a dataset, including over 10,000 cities worldwide and four types of urban data (DEM, water area, NTL, and built-up area).

\end{itemize}

\section{BACKGROUND AND RELATED WORK }
\subsection{Urban Morphology Generation}
Various factors shape urban morphology, including geographical situations, initial settlements, technological developments, economic activities, and others. Among them, physical geography (e.g., terrains and water bodies) and socioeconomic activity (e.g., nighttime lights) are important indicators.

Obviously, the terrains and water impact the cities' morphology \cite{CitySeedsGeographyOriginsEuropean2017-Bosker}, yet using them to simulate cities is very difficult. For example, terrains have many derivatives, such as the slope, orientation, and ruggedness (the degree to which the terrain is uneven). But as yet, it is not known to what extend these features could affect cities. This is also true for water areas, which are usually used as masks in urban morphology simulation but are not considered as a driving factor of cities. 

Compared with terrains and water, NTL data is a more widely used proxy to extract urban areas \cite{UrbanMappingUsingDMSPOLS2017-Li,QuantifyingUrbanAreasMultisourceData2020-Cao}. Two main NTL data sources are DMSP-OLS (Defense Meteorological Satellite Program's Operational Linescan System) \cite{MAPPINGCITYLIGHTSNIGHTTIMEDATA1997-Elvidge} and NPP-VIIRS (Suomi National Polar-orbiting Partnership satellite with the Visible Infrared Imaging Radiometer Suite sensor) \cite{elvidge2017viirs}. Based on NTL, various methods have been proposed to extract cities, including the threshold method \cite{QuantifyingUrbanAreasMultisourceData2020-Cao} and the Random Forest model \cite{huang2016mapping}. These methods classify each pixel as an urban area or a non-urban area independently. In addition, the classification-based methods can only extract the rough shape without the detailed structure of a city \cite{UrbanMappingUsingDMSPOLS2017-Li}.

\subsection{GANs Applied in Geospatial Domain}
GAN has been demonstrated to be a powerful tool to fit high-dimensional complex distributions \cite{GenerativeAdversarialNets2014-Goodfellow}. This advantage makes GAN good at modeling complex geospatial data with ubiquitous spatial effects (e.g., spatial dependence and heterogeneity). Zhu \emph{et al.} proposed CEDGANs to interpolate spatial data in geographical context, considering the spatial dependency effects \cite{SpatialInterpolationUsingConditionalGenerative2020-Zhua} . Ravuri \emph{et al.} combined GAN and RNN to incorporate both spatial and temporal dependencies \cite{ravuriSkilfulPrecipitationNowcasting2021}.

GAN has also been used to generate cities. Albert \emph{et al.} proposed a GAN model to learn urban morphology and generate fake city images that are both visually and statistically like real cities \cite{ModelingUrbanizationPatternsGenerativeAdversarial2018-Alberta}. Furthermore, they use conditional GAN to synthesize urban built-up maps using population distribution map, nightlight image, and water area map \cite{SpatialSensitivityAnalysisUrbanLand2019-Albert}. However, their model is limited by the sparse data and instability problems.

The above-mentioned GAN applications give a consistent body of evidence that GAN is good at learning the complex mapping between different geographical domains and considering spatial effects and structural connections in geo-spatial systems.

\begin{figure*}[h]
    \centering
    \includegraphics[width=\linewidth]{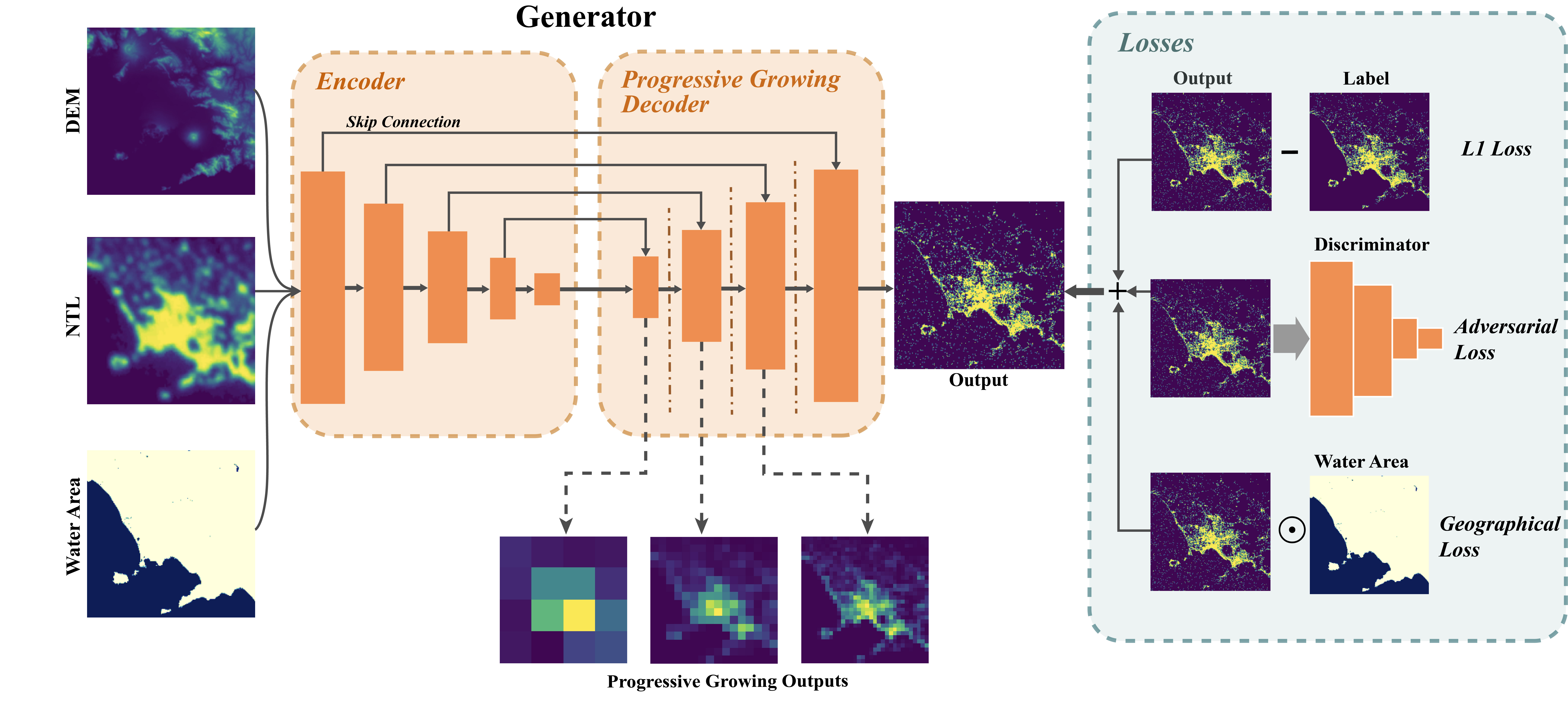}
    \caption{Architecture of MetroGAN}
    \label{fig:architecture}
\end{figure*}
\section{Methodology}
\subsection{Problem Formulation}
To introduce the urban simulation task to GANs, we represent the urban morphology as distribution maps and formalize the task as domain transfer of images. Specifically, transfer images from the source domain, the various aspects of urban attributes, to the target domain, the urban morphology. For the images of cities, we clipped them representing 100km $\times$ 100km spatial extent and resized each image to 256$\times$ 256 or 128$\times$ 128. Each pixel corresponds to around 390m $\times$ 390 m or 780m $\times$780m geographical extent.

The input images can be formulated as $\mathbf{X}=\{\mathbf{x}_1,\cdots,\mathbf{x}_n\}$, where $\mathbf{x}_i \in \mathbb{R}^{N\times N}$ is an image of source domains with size $N\times N$. In this work, the sources of $x_i$ can be DEM, NTL, and water area maps. The output is defined as $\mathbf{s} \in \mathbb{R}^{N \times N}$, which is a binary urban map (1 and 0 represent urban and non-urban areas respectively), and the goal of our model is to learn the conditional distribution $p(\mathbf{s} | \mathbf{X})$. The challenge of the learning process is twofold: how to capture the spatial effects and how to incorporate the geographical constraints.

\subsection{Geographical Domain Transfer with MetroGAN}
In general, a GAN consists of two networks, a generator (G) and a discriminator (D). D is trained to distinguish real images and fake images, and G is trained to confuse the discriminator. G and D compete with each other and finally reach an optimal equilibrium. Conditional GAN (CGAN) \cite{isolaImagetoImageTranslationConditional2017} applies GANs in the conditional setting and makes the generation process of G more controllable to solve complex generation tasks. Based on CGAN, pix2pixGAN \cite{isolaImagetoImageTranslationConditional2017} introduced GAN for image translation problems, which is a genetic approach to problems that traditional image translation tasks require specifically designed loss formulations. For our proposed model, the inputs of G are $\{{x_A},z\}$, where ${ x_A}$ is an image in domain A, and $z$ is a random vector in latent space, and the inputs of D are $\{x_A,y\}$ and $\{x_A,G(x_A,z)\}$, where $y$ is the label image in target domain and $G(x_A,z)$ is the output image of G. D not only distinguishes whether the input is a "real" image, but also distinguishes whether the input pair matches. The optimization objective of the original pix2pixGAN is
\begin{equation}
    G^*=\arg \min_G \max_D {\mathcal {L}}_{cGAN}(G,D)+\lambda {\mathcal{L}}_{L1}(G).
\end{equation}
\begin{equation}
\begin{aligned}
    {\mathcal {L}}_{cGAN}(G,D)=&{\mathbb{E}}_{x,y}[logD(x,y)]+\\
    &{\mathbb{E}}_{x,z}[log(1-D(x,G(x,z)))],
\end{aligned}
\end{equation}
\begin{equation}
    {\mathcal{L}}_{L1}(G)={\mathbb{E}}_{x,y,z}[\Vert y-G(x,z) \Vert_1],
\end{equation}
where ${\mathcal {L}}_{cGAN}(G,D)$ is the objective of a conditional GAN and ${\mathcal{L}}_{L1}(G)$ is the L1 loss encouraging pixel-by-pixel accuracy and less blurring. $x$ is the input images in source domain, $y$ is the label images in target domain, and $z$ is a random vector sampled from normal distribution.

Here, we propose a novel GAN-based framework to solve the domain transfer tasks for urban images, namely MetroGAN. As shown in Figure \ref{fig:architecture}, we adopt L1 loss and the U-Net generator in pix2pixGAN \cite{isolaImagetoImageTranslationConditional2017} to reinforce the information of corresponding locations in input and output images. To incorporate the complicated hierarchical features of the urban system, we propose a progressive growing training procedure in our model. We also design a geographical constraint loss to impose a soft constraint that cities cannot be on the water. Besides, to solve the instability problem in the training process, we transform the adversarial loss to minimize the Pearson $\chi^2$ divergence following LSGAN \cite{LeastSquaresGenerativeAdversarialNetworks2017-Maoa}. Our final objective is
\begin{equation}
\begin{aligned}
    G^*=\arg \min_G &\max_D {\mathcal {L}}_{LSGAN}(G,D)+\\&\lambda_{L1} {\mathcal{L}}_{L1}(G)+\lambda_{geo} \mathcal{L}_{geo}(G),
\end{aligned}
\end{equation}
where ${\mathcal {L}}_{LSGAN}(G,D)$ is the least square adversarial loss \cite{LeastSquaresGenerativeAdversarialNetworks2017-Maoa} replacing ${\mathcal {L}}_{cGAN}(G,D)$ (see Section 3.4), and  ${\mathcal{L}}_{geo}(G)$ is the geographical constraint loss of water area (see Section 3.3). $\lambda_{L1}$ and $\lambda_{geo}$ are the scale factors of L1 loss and geographical constraint loss, which are set as 50, 100 in our experiments. 

\subsection{Geographical Constraint of Water Body}
Although the deep generative model can naturally learn the impact of the physical geography,  physics-informed machine learning enlightens that we can add some geographical constraints to the loss function, enforcing corresponding learning biases to the model \cite{PhysicsinformedMachineLearning2021-Karniadakis}. With more learning biases, the model can learn the mappings in line with geographical knowledge and own better performance.

The geographical constraint that the buildings cannot be located on water can be easily implemented. We use the Hadamard product of the water area and the generated city image as the loss of the water constraint. The water area image is a binary image, in which 1 represents water and 0 represents land. The output of MetroGAN is a one-channel image, in which the value of the pixel represents the probability of being an urban area. Their Hadamard product can filter the pixels that generate urban area on the water, and the larger the probability of being an urban area, the larger the loss is. The geographical constraint loss can be expressed as follows:
\begin{equation}
    \mathcal{L}_{water}(G)=-\mathbb{E}_{x,z}[x_{water}\odot G(x,z)]
\end{equation}
where $x_{water}$ is the water area image.

\subsection{Stabilizing the Training Process}
One challenge in the study of generative adversarial networks is the instability of its training \cite{SpectralNormalizationGenerativeAdversarialNetworks2018-Miyato}. This problem is very serious in the setting of urban morphology simulation, which is possibly due to the fact that the built-up area is a binary image, and the dark pixels (value = 0) account for a large proportion in most cities. 

Widely applied methods that stabilize the training process include Wasserstein GAN with Gradient Penalty (WGAN-GP) \cite{ImprovedTrainingWassersteinGANs2017-Gulrajania}, SpectralNorm \cite{SpectralNormalizationGenerativeAdversarialNetworks2018-Miyato}, and Least Squares GAN (LS-GAN) \cite{LeastSquaresGenerativeAdversarialNetworks2017-Maoa}.
WGAN-GP is an improved version of original Wasserstein GAN \cite{WassersteinGAN2017-Arjovsky}. It transforms the objective function from KL divergence to Wasserstein distance and imposes a gradient penalty to softly satisfy the Lipschitz constraint that the Wasserstein distance required. SpectralNorm imposes weight normalization using spectral coefficient, and LS-GAN transforms the objective function to minimize the Pearson $\chi^2$ divergence. 

Our results show that WGAN-GP does not work well in image translation tasks, while SpectralNorm and LS-GAN are both effective. For a detailed comparison, the results of the two methods are listed in Section 5.6. We found that LS-GAN can not only stabilize the training, but also improve the performance of the model. Thus, the least square loss is finally chosen as our adversarial loss. The objective of LSGAN is:
\begin{equation}
    \begin{aligned}
    \mathcal{L}_{LSGAN}(G,D) = & \frac{1}{2} \mathbb{E}_{x,y}\left[(D({x,y})-1)^{2}\right]+\\&\frac{1}{2} \mathbb{E}_{x,z}\left[(D(x,G({x,z})))^{2}\right] \\
    \end{aligned}
\end{equation}

\subsection{Progressive Growing Structure}
To further improve the model, we introduce the progressive training structure, which has been validated to stabilize the model and produce high-quality outputs \cite{karras2017progressive}. The progressive training procedure starts from low-resolution images and then gradually adds layers to deal with high-resolution images. Correspondingly, an urban system is a highly self-organized system with self-similarity and hierarchical structure, and with progressive training, our model can incorporate these characteristics better. 

Based on the thought of progressive growth, we design our generator (U-Net) and discriminator to add layers progressively in the learning process. For the generator, we split it as the encoder block and the progressive growing decoder block. The structure of the encoder block is fixed, and the decoder block grows from producing 4$\times$4 images to producing 256$\times$256 images. Similar to the decoder block, the discriminator’s layers are also progressively added in synchrony with the decoder, from taking 4$\times$4 images as input to taking 256$\times$256 images. For resolution transition, we design a mechanism based on \cite{karras2017progressive} to fade the new layers in and preserve well-trained last-level parameters. After adding new layers to the decoder block, we weighted sum the output of the new layers with weight $\alpha$ and upsampled outputs of the last layer with weight $(1-\alpha)$ as the output. The weight $\alpha$ gradually increases to 1 to fade the new layer in smoothly. The procedure of the discriminator is a mirror with the decoder, summing the output of the new layer and downsampled results of input as the new input of D. Section 5.5 describes the improvement brought by the progressive training methodology.

\section{Evaluation Framework}
To evaluate our results and compare the similarity between the generated city and the labeled city, we need to consider two questions: 
\begin{enumerate}
    \item How much does the generated city image look like the labeled city?
    \item Does the generated city have similar complex features as real cities?
\end{enumerate}

In response to the first question, we compare our model with existing models at three levels: the pixel level, the multi-scale spatial level, and the perceptual level, focusing on the quantitative differences between the cities generated by different models and the labeled cities. For the second question, complexity science sheds some light. As a typical complex system, cities follow some macroscopic laws, such as the Zipf's law \cite{ZipfLawAllNaturalCities2011-Jiang,ModellingUrbanGrowthPatterns1995-Makse} and the fractal structure \cite{batty1994fractal}. Thus, we use the fractal dimension and Zipf's exponent to evaluate the generated cities at a macroscopic level. Next, we detail the evaluation metrics under different levels (Table \ref{tab:evaluation framework}).  

\begin{table}[h]
\centering
\caption{Summary of the evaluation framework}
\label{tab:evaluation framework}
\begin{tabular}{ccc} 
\toprule
\textbf{Purpose}            & \textbf{Level}      & \textbf{Metrics}                                                         \\ 
\midrule
\multirow{3}{*}{Comparison} & Pixel Level         & PSNR                                                                     \\
                            & multi-scale spatial level       & SPM                                                                      \\
                            & Perceptual Level    & SSIM / LPIPS                                                             \\
\midrule
Validation                  & ~Macroscopic Level~ & \begin{tabular}[c]{@{}c@{}}Fractal Dimension /\\Zipf's Law\end{tabular}  \\ 
\bottomrule
\end{tabular}
\end{table}

\subsection{Pixel Level}
At the pixel level, we use MSE (mean square error)  and PSNR (peak signal to noise ratio) as the pixel-by-pixel accuracy evaluation metrics. 
\begin{equation}
MSE=\frac{1}{mn}\sum_{i=0}^{m-1}\sum_{j=0}^{n-1}\Vert I(x,y)-K(x,y)\Vert^2
\end{equation}
\begin{equation}
PSNR=10\log_{10}(\frac{MAX_I^2}{MSE})
\end{equation}
            
\subsection{Multi-Scale Spatial Level}
Previous studies dealing with spatial data usually represent the location information using the hierarchical-resolution gridding \cite{ProgRPGANProgressiveGANRoutePlanning2021-Fu,GeographyAwareSequentialLocationRecommendation2020-Lian}, (e.g., locate the objects in grid segmentation of 4$\times$4, 8$\times$8, …). Following this idea, we improve the spatial pyramid matching (SPM) metric \cite{BagsFeaturesSpatialPyramidMatching2006-Lazebnik} at this level. SPM is a solution to evaluate the similarity of the distribution of feature points considering spatial information, and we assume that every urban area pixel is a feature point to calculate the SPM score. The procedure of calculating the SPM score can be divided into the following steps:

First, for images X and Y, divide them with different scales (in $l$ layer divide the side length into $2 ^ L$). Under the L-layer segmentation, the feature points in the corresponding sub-region of two images are counted for all subregions. The smaller count is taken as the value of each subregion. Then sum all subregions as the layer's score ${\mathcal I}(H_X^l,H_Y^l)$.
\begin{equation}
    {\mathcal I}(H_X^l,H_Y^l)=\sum_{i=1}^{D}min(H_X^l(i),H_Y^l(i))
\end{equation}

The scores for each layer are summed as follows (The finer scale, the larger weights):
\begin{equation}
{\mathcal K}^L(X,Y)={\mathcal I}^L+\sum_{ l=1}^{L-1}\frac{1}{2^{L-l}}(\mathcal{I}^l-\mathcal{I}^{l+1})
\end{equation}

To punish the behavior of generating much larger area than the label to cover as many label pixels as possible, we regularize the score of SPM by the max count of bright pixels in output and label images. Besides, because the magnitude of regularized SPM score is so small, we multiply it by 100.

\begin{equation}
SPM=\frac{1}{\max(N(label),N(output))}{\mathcal K}^L(X,Y)
\end{equation}

\subsection{Perceptual Level}
At the perceptual level, we use SSIM (Structural Similarity Index) and LPIPS ( Learned Perceptual Image Patch Similarity) \cite{UnreasonableEffectivenessDeepFeaturesPerceptual2018-Zhanga} metrics to extract images' structural features. SSIM can only extract shallow attributes, while LPIPS, with the deep neural network, can extract deep structural features of an image. 

\subsection{Macroscopic Level}
The fractal feature is one of the most important morphological characteristics of the city and the fractal dimension is proposed to measure this feature and the degree of self-similarity. Here, we use the box counting \cite{sarkar1994efficient} method to calculate the fractal dimension, which is defined as
\begin{equation}
    dim_{box}(S):=\lim_{\epsilon \to 0}\frac{logN(\epsilon)}{log(1/\epsilon)}, 
\end{equation}
where $N(\epsilon)$is the count of boxes needed to cover the city boundaries using the box with side length $\epsilon$. 

Unlike the fractal dimension, Zipf's law is a statistical law that describes the size distribution of the settlements in the urban system. According to Zipf's law, the relationship between the size of the settlements and their frequencies of as follows: 
\begin{equation}
    P(S)=kS^{\gamma},
    \label{equ:zipf}
\end{equation}
where $\gamma$ is approximately -2. \cite{ZipfLawAllNaturalCities2011-Jiang}. 

\section{Experiments}

\subsection{Data}
\begin{table}[t]
\centering
\caption{Summary of data sources}
\label{tab:datasource}
\resizebox{\linewidth}{!}{%
\begin{tabular}{lllll} 
\toprule
\textbf{Type}                                                 & \textbf{Source} & \begin{tabular}[c]{@{}l@{}}\textbf{Spatial}\\\textbf{Resolution}\end{tabular} & \begin{tabular}[c]{@{}l@{}}\textbf{Temporal}\\\textbf{Resolution}\end{tabular} & \begin{tabular}[c]{@{}l@{}}\textbf{Temopral}\\\textbf{Range}\end{tabular}  \\ 
\midrule
\begin{tabular}[c]{@{}l@{}}Administrative\\Area\end{tabular} & GADM            & None                                                                            & None                                                                             & None                                                                        \\ 
\midrule
Built-up Area                                                 & GHSL            & 30m                                                                             & 15yr                                                                             & 1975-2014                                                                   \\ 
\midrule
Population                                                    & LandScan        & 1km                                                                             & 1yr                                                                              & 2000-2020                                                                   \\ 
\midrule
DEM                                                           & SRTM            & 30m                                                                             & None                                                                             & None                                                                        \\ 
\midrule
Water Area                                                    & GSW             & 30m                                                                              & None                                                                             & None                                                                        \\ 
\midrule
\multirow{2}{*}{Nighttime light}                                   & DMSP-OLS        & 1km                                                                             & 1yr                                                                              & 1992-2014                                                                   \\ 
\cmidrule{2-5}
                                                              & NPP-VIIRS       & 500m                                                                            & 1yr                                                                              & 2014-2020                                                                   \\
\bottomrule
\end{tabular}
}
\end{table}
\begin{figure}[b]
    \centering
    \includegraphics[width=\linewidth]{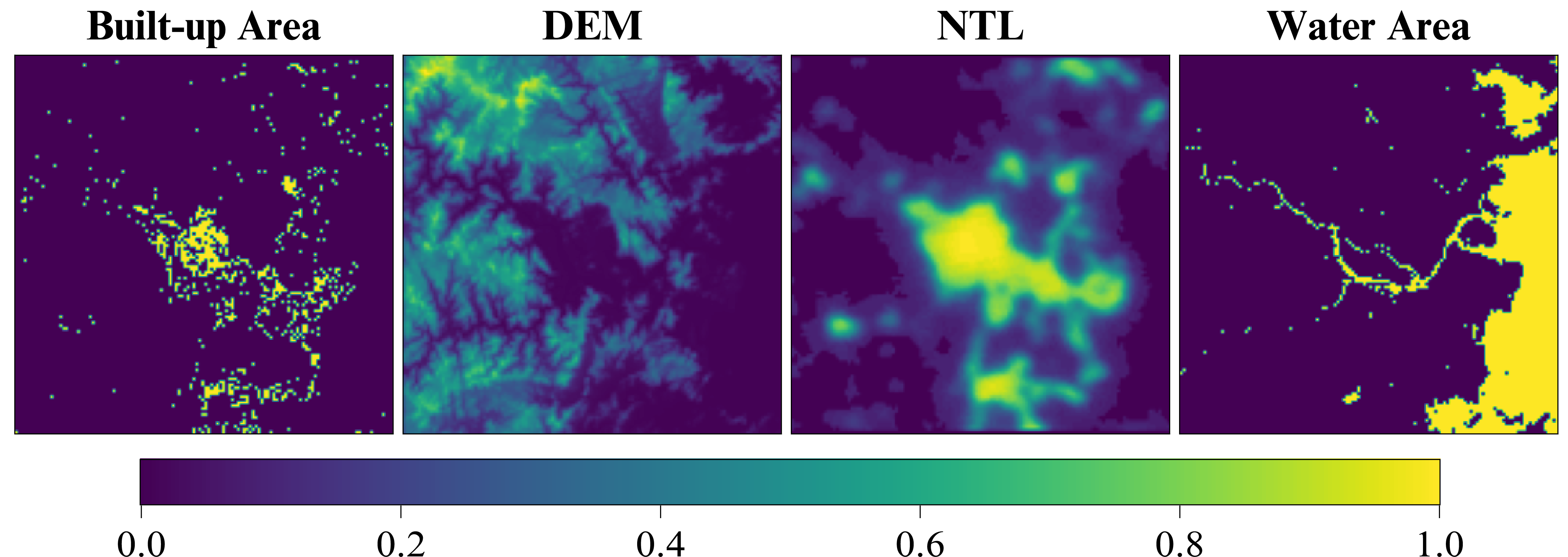}
    \caption{Input samples. The built-up area and water area are binary images, in which value 1 means urban/water area. The value of DEM and NTL are scaled to [0,1].}
    \label{fig:input_sample}
\end{figure}
To perform our experiments, we construct and publish a dataset of global cities. Table \ref{tab:datasource} presents the types, sources, spatial resolution, temporal resolution, and temporal range of our dataset. The GHSL built-up area data \cite{GHSDUCR2019AGHSDegree2021-Schiavina}, published in 1975, 1990, 2000 and 2014, are binary images that urban area pixels have a value of 1 and non-urban area pixels are 0. Each pixel of the DEM data \cite{ShuttleRadarTopographyMissionSRTM2017-EarthResourcesObservationAndScienceEROSCenter} is the average elevation of the corresponding area. The water area data \cite{HighresolutionMappingGlobalSurfaceWater2016-Pekel} is also a binary map that the water area pixels have a value of 1, and others are set to 0. We use two types of nightlight datasets: DMSP-OLS \cite{MAPPINGCITYLIGHTSNIGHTTIMEDATA1997-Elvidge} and NPP-VIIRS \cite{elvidge2017viirs}. NPP-VIIRS NTL data has improved the problems of blurring and saturation of the DMSP-OLS data to a great extent. However, its temporal range is relatively short and cannot be used to extract historical urban areas before 2012.

We select global cities with more than 10,000 populations (see Appendix for details). Then we clip city images representing 100km $\times$ 100 km spatial extent and resize the images to 128$\times$128 (for DMSP-OLS data) and to 256$\times$256 (for NPP-VIIRS data). After filtering the built-up area images with less than 1\% amount of signal (bright pixels account for less than 1\%), only approximately 6,000 cities were left. To solve the problem of insufficient training data, we combine the data of 2000 and 2014 and obtain a dataset with 11,873 cities. Notably, the NPP-VIIRS NTL data are published from 2012 to 2021, and only one snapshot of the built-up area data (2014) is within this period. We therefore construct the 256$\times$256 dataset with single-year data and 128$\times$128 dataset with multi-year data.

\begin{figure*}[htbp]
    \centering
    
    \begin{subfigure}[b]{0.35\linewidth}
        \centering
        \includegraphics[height=2.9in]{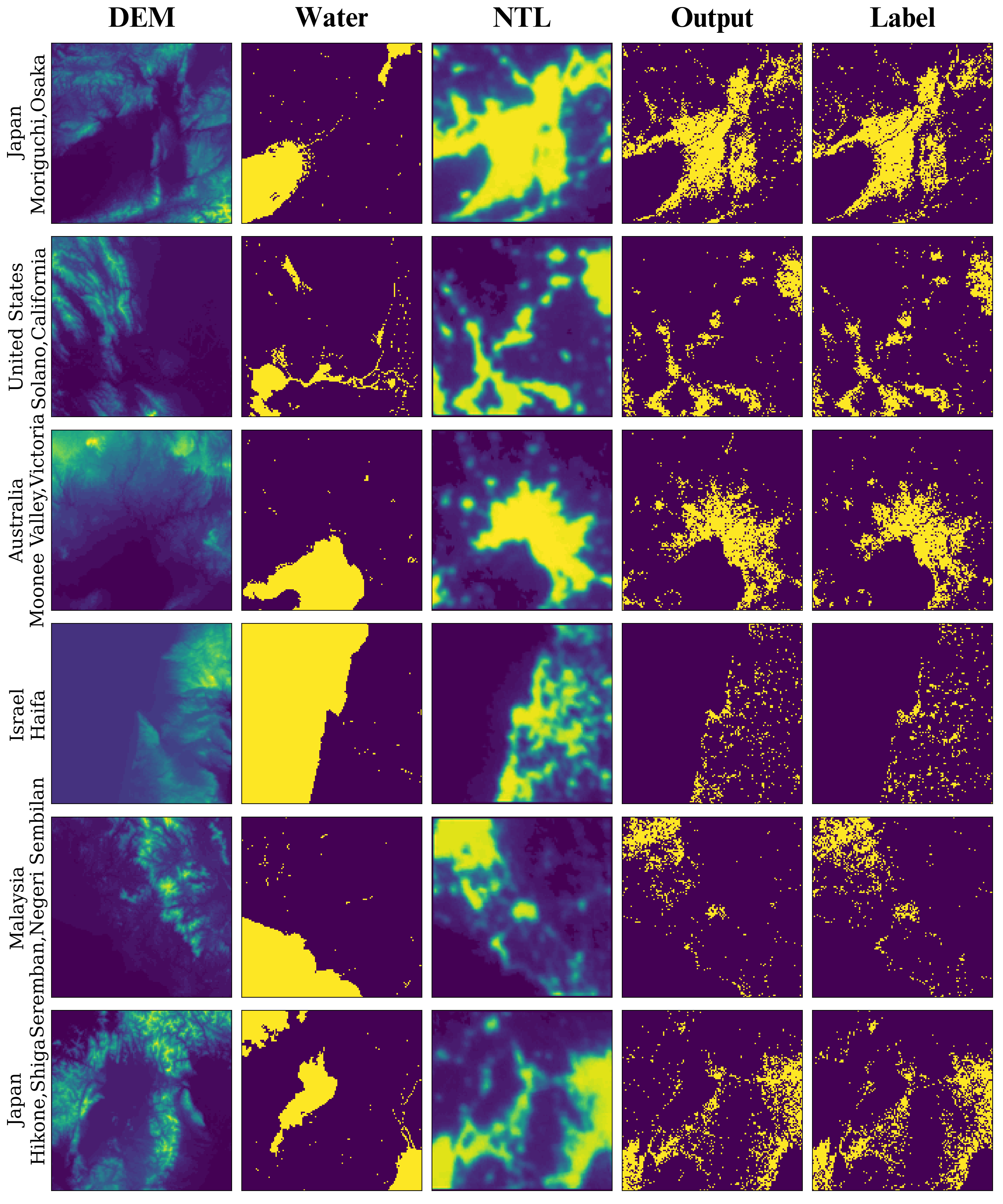}
        \subcaption{Results of 128$\times$128 Dataset}
        \label{fig:128 results}
    \end{subfigure}\hfill
    \begin{subfigure}[b]{0.35\linewidth}
        \centering
        \includegraphics[height=2.9in]{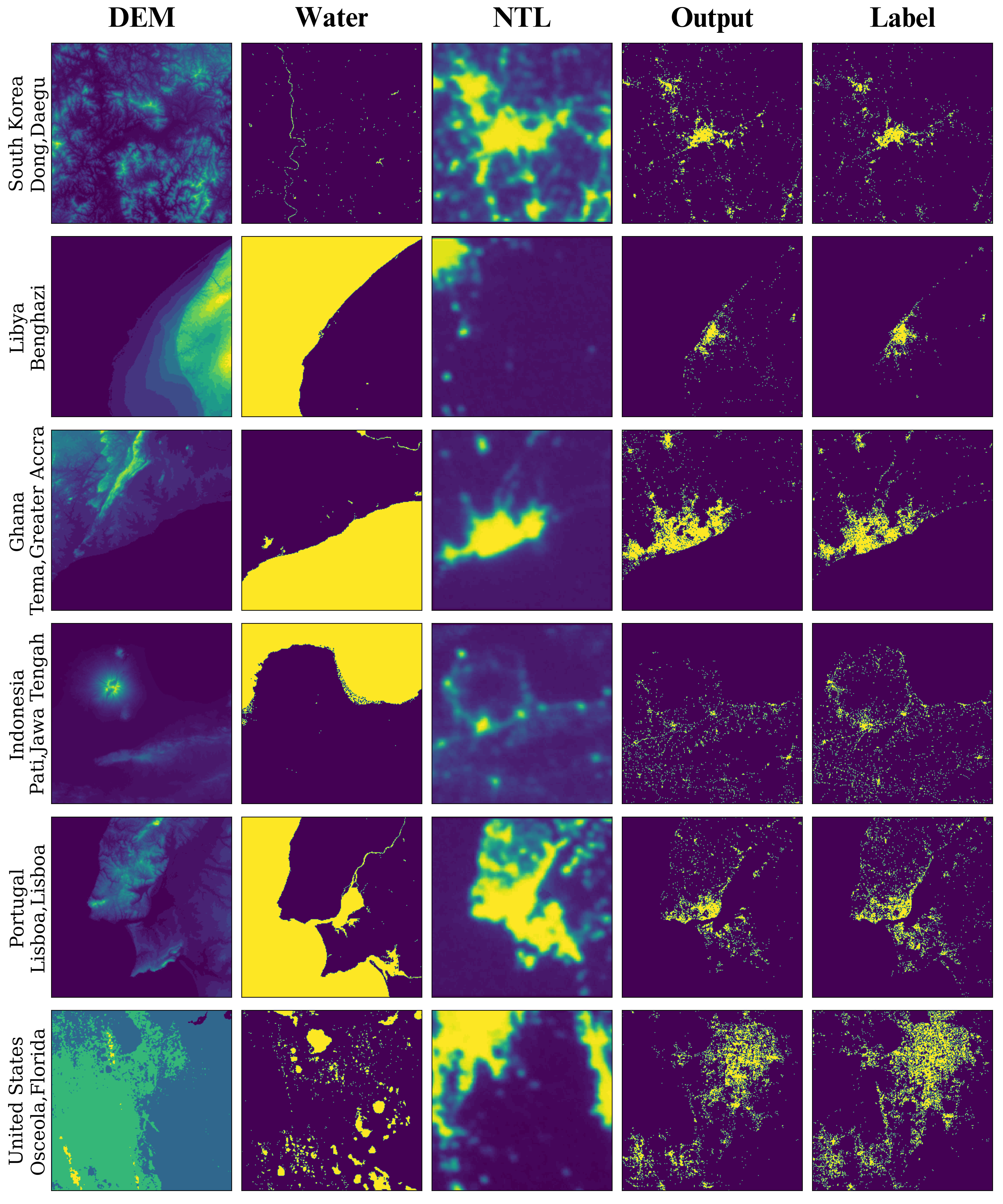}
        \subcaption{Results of 256$\times$256 Dataset}
        \label{fig:256 results}
    \end{subfigure}
    \begin{subfigure}[b]{0.28\linewidth}
        \centering
        \includegraphics[height=2.9in]{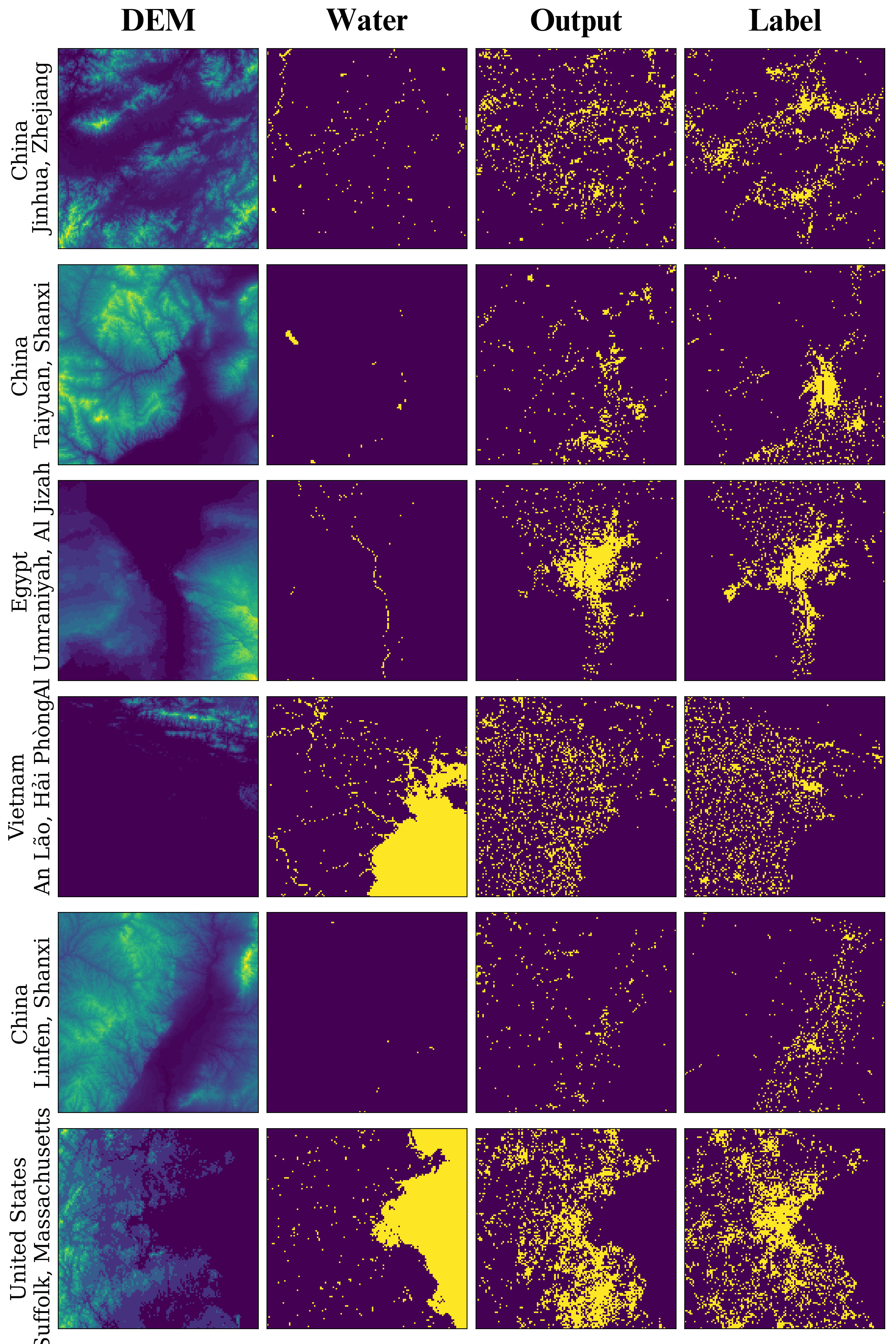}
        \subcaption{Results of physical geography inputs}
        \label{fig:results_of_nature}
    \end{subfigure}
    \caption{Randomly sampled results of three experiments}
    \label{fig:results}
\end{figure*}

\subsection{Baselines}
To show the effectiveness of our model, we compare MetroGAN with the following baselines:
\begin{itemize}[leftmargin=*]
    \item XGBoost \cite{chen2016xgboost} is a scalable tree boosting system based on gradient boosting tree model. Methods based on classification have been widely used in geography to simulate urban area and the XGBoost classification model is a state-of-the-art method of them \cite{UrbanMappingUsingDMSPOLS2017-Li} . We input each pixel's information (NTL, DEM and water area) to the model to classify a pixel as urban area or not.
    \item U-Net \cite{UNetConvolutionalNetworksBiomedicalImage2015-Ronneberger} is a popular deep learning method for image translation tasks, but has not been applied in urban simulation tasks. Using U-Net, we apply DEM, water, and NTL as inputs to generate cities.

    \item CityGAN \cite{SpatialSensitivityAnalysisUrbanLand2019-Albert} (we use the name of related GitHub repository for the model) is the state-of-the-art GAN-based model for the urban morphology simulation task. This model is based on pix2pix GAN and takes DEM, water, and NTL as inputs in our experiments.
\end{itemize}

\subsection{Comparison with Baselines}
\begin{figure}[htpb]
    \centering
    \includegraphics[width=\linewidth]{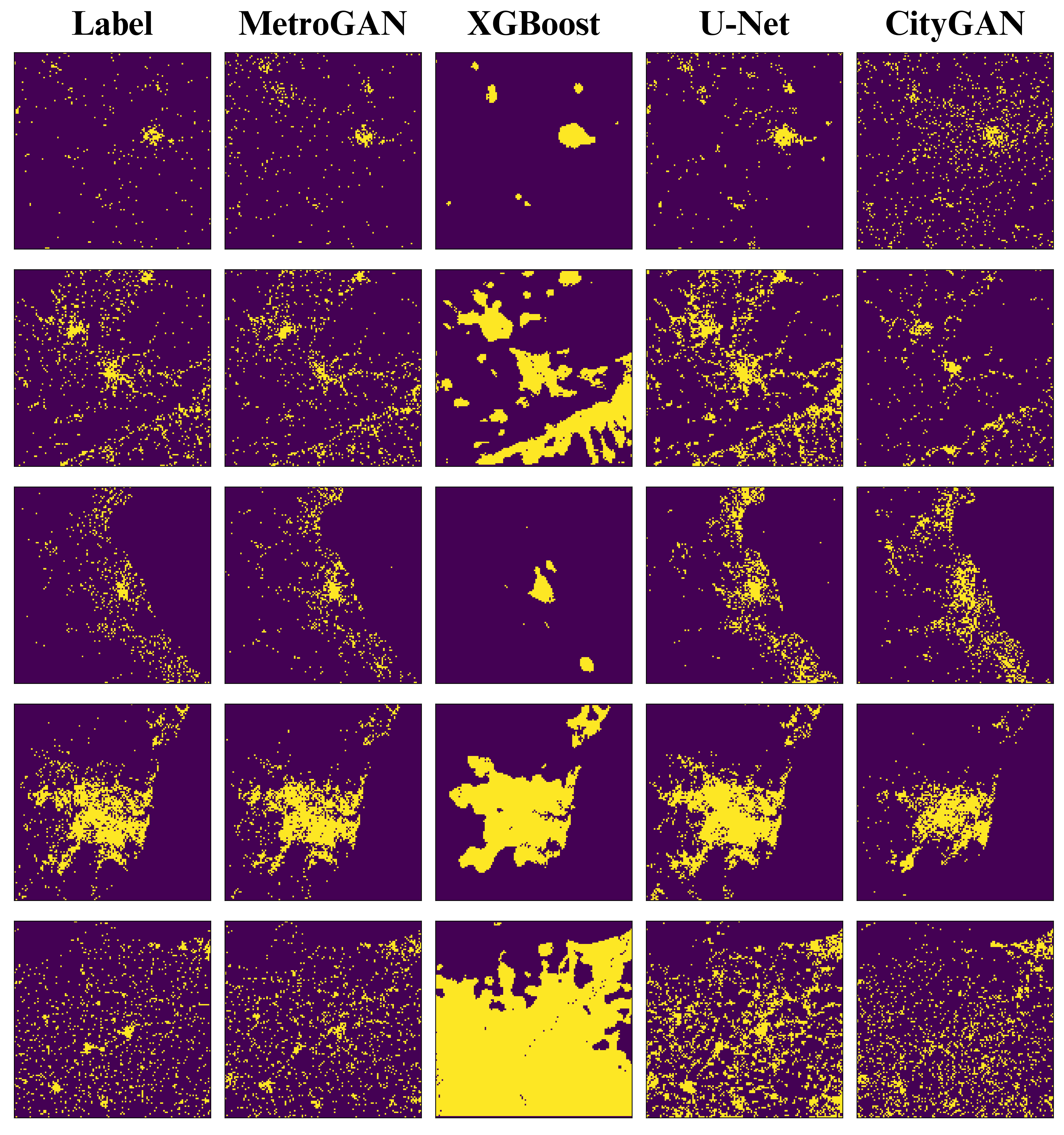}
    \caption{Comparison of results of MetroGAN and baselines}
    \label{fig:comparison}
\end{figure}

Some representative images of the results are shown in Figure \ref{fig:results} and the results of MetroGAN and baselines are summarized in Table \ref{tab:results}. To visually show the comparison, we also present the ground truth and the generated urban area of different models in Figure \ref{fig:comparison}. 


Table \ref{tab:results} shows that MetroGAN outperforms all other baselines in both 128$\times$128 dataset and 256$\times$256 dataset. Due to the blur of nightlight data, it is more challenging to produce cities using the 128$\times$128 dataset. In this dataset, our model achieves 25.3\% and 21.1\%  improvements over the best-performing baseline in terms of SPM and LPIPS, respectively, indicating that our model is good at capturing the spatial and deep structured information of the urban system. Our model also improves 7.1\% performance in PSNR and 6.5\% in SSIM compared with U-Net. Compared with CityGAN, the improvements are up to 20.39\% and 19.8\%. This reveals that our model comprehensively improves the performance in all aspects, especially when facing blurring inputs. Interestingly, in the 256$\times$ 256 dataset, the performance in PSNR and SSIM of the four methods are much closer than in the 128$\times$128 dataset (5.6\% in PSNR and 0.0\% in SSIM), but our model can still outperform the best baseline in terms of SPM and LPIPS (13.8\% in SPM and 5.2\% in LPIPS). This implies that with the increase of the spatial resolution of NTL maps, all models own a better performance on pixel-by-pixel accuracy, but generating the overall spatial structures and leveraging the information of terrains and water body is still hard for baselines, especially XGBoost.
\begin{table}[H]
\centering
\caption{Performance comparison}
\label{tab:results}
\resizebox{0,98\linewidth}{!}{%
\begin{tabular}{ccccc} 
\toprule
         & \textbf{Pixel Level} & \begin{tabular}[c]{@{}c@{}}\textbf{Multi-scale}\\\textbf{Spatial Level}\end{tabular} & \multicolumn{2}{c}{\textbf{Perceptual Level}}  \\ 
\cmidrule(lr){2-2}\cmidrule(lr){3-3}\cmidrule(l){4-5}
         & PSNR$\uparrow$       & SPM$\uparrow$                                                                        & SSIM$\uparrow$  & LPIPS$\downarrow$            \\ 
\midrule
\multicolumn{5}{l}{\textbf{128$\times$128 Dataset }}                                                                                                                    \\ 
\midrule
XGBoost  & 9.3806              & 0.1577                                                                               & 0.4659          & 0.6593                       \\
U-NET    & 11.7633              & 0.1772                                                                               & 0.4771          & 0.3405                       \\
CityGAN  & 10.4691              & 0.1669                                                                               & 0.4241          & 0.3408                       \\
MetroGAN & \textbf{12.6011}     & \textbf{0.2232}                                                                      & \textbf{0.5083} & \textbf{0.2687}              \\ 
\midrule
\multicolumn{5}{l}{\textbf{256$\times$256 Dataset }}                                                                                                                    \\ 
\midrule
XGBoost  & 15.6972              & 0.2122                                                                               & 0.5839          & 0.4463                       \\
U-NET    & 15.5728              & 0.2201                                                                               & 0.6201          & 0.4110                       \\
CityGAN  & 15.4984              & 0.2239                                                                               & 0.5408          & 0.4053                       \\
MetroGAN & \textbf{16.5702}     & \textbf{0.2547}                                                                      & \textbf{0.6203} & \textbf{0.3842}              \\
\bottomrule
\end{tabular}
}
\end{table}

\subsection{Validation of Macrocopic Level Metrics}
\begin{figure}[t]
    \centering
    \includegraphics[width=\linewidth]{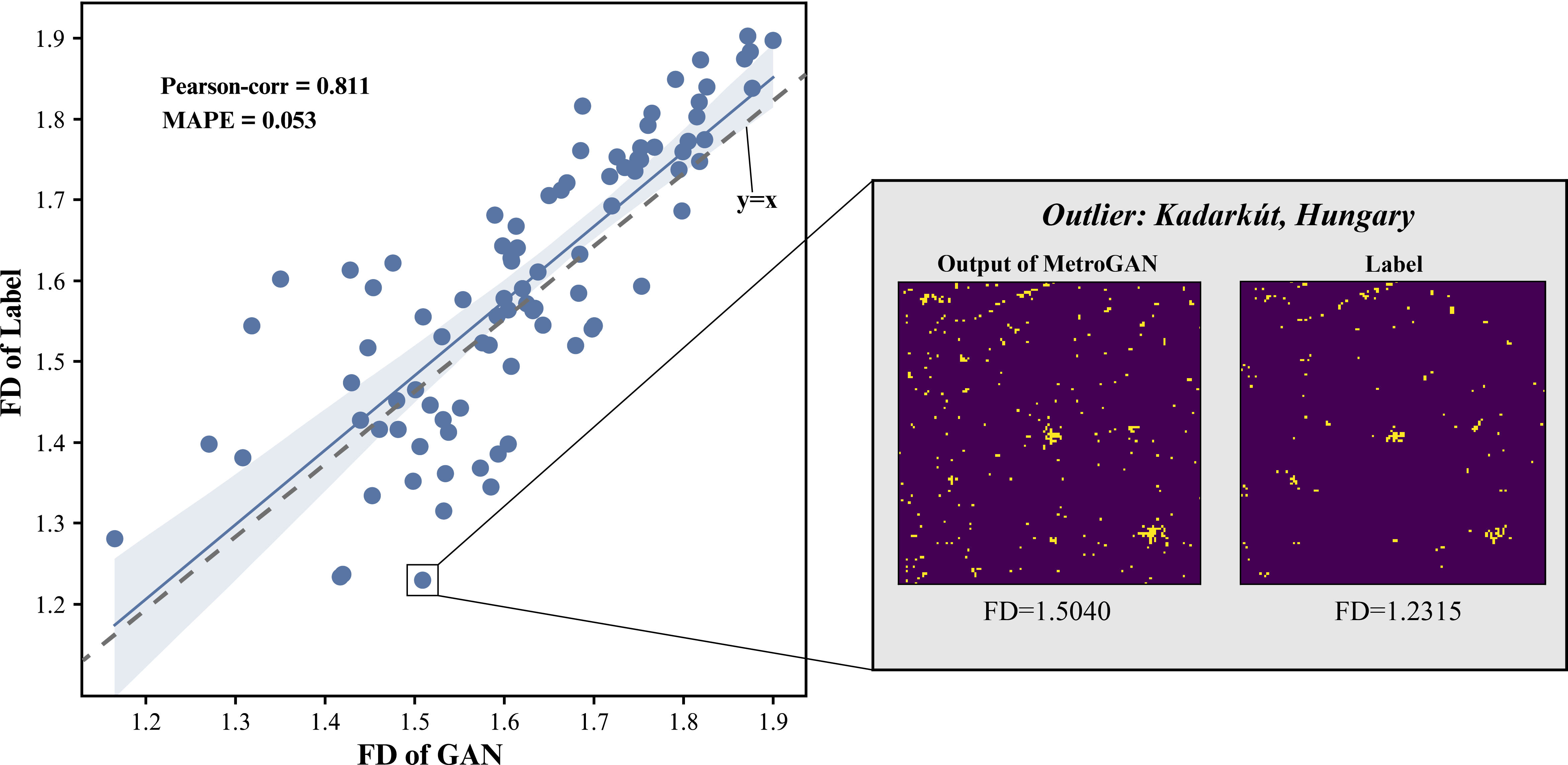}
    \caption{Results of fractal dimension}
    \label{fig:results_of_fractaldimension}
\end{figure}

From the macroscopic level, we validate whether the generated images are like a real city by calculating the Zipf's exponent and fractal dimensions. Specifically, we calculated the fractal dimension of all generated cities (x-axis
 of Figure \ref{fig:results_of_fractaldimension}) and label cities (y-axis
 of Figure \ref{fig:results_of_fractaldimension}) in the test set. The Pearson correlation coefficient between fractal dimensions of the label cities and the generated cities is 0.823 and the MAPE is 0.052. These results reveal that the fractal dimensions of generated cities are very close to the label cities, indicating that our model perfectly captures the cities' characteristics of fractality. For those outliers, we find that most of them are cities whose built-up images contain few pixels (One typical outlier is shown in Figure \ref{fig:results_of_fractaldimension} and others are shown in appendices). The small amount of urban area might mean the urban system is not mature, and the calculation of fractal dimension has low sensitivity; that is, few wrongly classified bright pixels can affect the fractal dimension to a large extent.

\begin{figure}[h]
    \centering
    \includegraphics[width=0.99\linewidth]{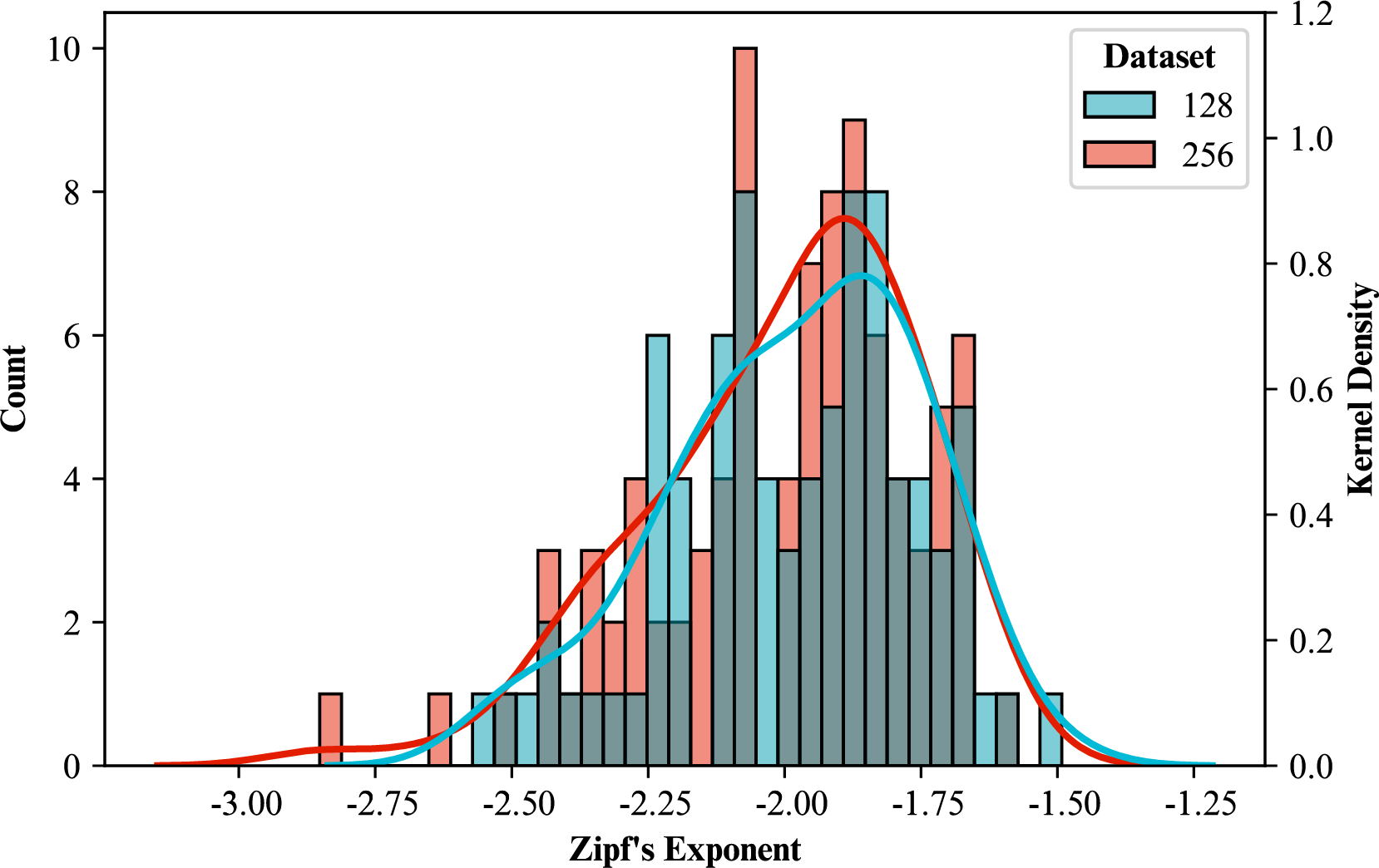}
    \caption{Distribution of Zipf's exponents for 128$\times$128 dataset and 256$\times$256 dataset}
    \label{fig:zipf_distribution}
\end{figure}

Then, we calculate the Zipf's exponent by log-transforming Eq. \ref{equ:zipf} to $ln(P(A))=\gamma ln(A)+C$. The exponent $\gamma$ is the coefficient of the linear regression model. Figure \ref{fig:zipf_distribution} shows that the $\gamma$s of our generated cities are around -2, indicating a well-fitted Zipf's law. Specifically, in the 256$\times$256 dataset, the exponents fall within [-1.75,-2.25] in 76/100 cities and within [-1.5,-2.5] in 97/100 cities. In 128$\times$128 dataset, the coefficients fall within [-1.75,-2.25] in 84/100 cities and within [-1.5,-2.5] in 98/100 cities.
\subsection{Generating Urban Built-up Area from Physical Geographic Features}
We also conducted an experiment to simulate urban areas just from DEM and water using the 128$\times$128 dataset. This experiment can be of great use in the research of urban geography to shed some lights on key questions, such as to what extent the physical geography can influence the urban form or what kind of physical geographic variables influence the urban form most. Some samples of the results are shown in Figure \ref{fig:results_of_nature}. 

The results show that in some special natural conditions, such as the city being along a valley, in a basin, or by the bay, our model can generate a rough shape and some textures of the city. This implies that when the blooming of the cities is restricted by spatial conditions, the impact of nature is greatly enhanced. Conversely, if a city is in a vast plain and has no spatial constraint, the impact of nature is relatively low, which is revealed in our results that cities grow isotropically. In these cases, other types of urban data, such as the location of the first settlement and the distribution of natural resources, may be useful for further explorations.
\begin{table*}
\centering
\caption{Ablation analysis. The components that are used in MetroGAN and improve the performance are boldfaced.}
\label{tab:ablation}
\begin{tabular}{ccccccccccc} 
\toprule
\multicolumn{5}{c}{Ablation 1}                                                   &  & \multicolumn{5}{c}{Ablation 2}                                                    \\ 
\cmidrule{1-5}\cmidrule{7-11}
      & PSNR$\uparrow$   & SPM$\uparrow$   & SSIM$\uparrow$  & LPIPS$\downarrow$ &  &       & PSNR$\uparrow$   & SPM$\uparrow$   & SSIM$\uparrow$  & LPIPS$\downarrow$  \\ 
\cmidrule{1-5}\cmidrule{7-11}
Base1 & 10.4690          & 0.1669          & 0.3673          & 0.3408            &  & Base2 & 11.7631          & 0.2115          & 0.4792          & 0.2843             \\
WGP   & 10.5210          & 0.1671          & 0.4080          & 0.3441            &  & -DV   & 9.4430           & 0.1701          & 0.3112          & 0.3129             \\
SN    & 11.7227          & 0.1872          & 0.4661          & 0.3076            &  & +GC   & \textbf{12.2924} & \textbf{0.2227} & \textbf{0.5030} & \textbf{0.2725}    \\
LS    & \textbf{11.7631} & \textbf{0.2115} & \textbf{0.4792} & \textbf{0.2843}   &  & +PT   & \textbf{12.5146} & \textbf{0.2219} & \textbf{0.4938} & \textbf{0.2610}    \\
\bottomrule
\end{tabular}
\end{table*}
\subsection{Ablation Study}
To comprehensively analyze the effectiveness of the various components in our model, we conduct two ablation experiments. 
\subsubsection{The structures to stabilize training}
We test three methods to stabilize the training process and improve the performance. In Ablation 1, we use pix2pixGAN as our base model (Base1 in Table \ref{tab:ablation}) and add the following three structures in the base model to test their effectiveness.
\begin{itemize}[leftmargin=*]
    \item WGAN-GP (Wasserstein GAN with Gradient Penalty): We replace the original loss with Wasserstein loss and add the penalty term into the loss function. 
    \item LS (Least Square Loss): We use the mean square error loss, replacing the original binary cross entropy loss, when calculating the loss of the discriminator.
    \item SN (Spectral Normalization): We add spectral norm operation to all convolutional layers in the discriminator.
\end{itemize}

The results are summarized in Table \ref{tab:ablation}, Ablation1. From the results, we can tell WGAN-GP has little or no improvement in the base model. Meanwhile, we find that the non-convergence phenomena of the original model still appear sometimes. Spectral Normalization and Least Square Loss can eliminate the problem of non-convergence and unstable training, and the results show that using LS Loss can bring larger improvement than Spectral Normalization. In addition, it is worth mentioning that adding both LS and SN does not work because both of them will constrain the gradients to too small.

\subsubsection{The components to improve the performance}
To validate the other components in our model, we did another ablation experiment using the LSGAN trained on multi-year 128$\times$128 dataset as the base method. It should be mentioned that we also use the water maps as masks to outputs of the base method, constraining cities are not on the water. We consider the following variants:
\begin{itemize}[leftmargin=*]
    \item -DV (data volume): We remove the data of the year 2000 from the 128$\times$128 dataset, getting a single-year dataset with 6384 cities.
    \item +GC (Geographical constraint): We add our geographical constraint loss in addition to the least square loss and L1 loss.
    \item +PT (Progressive Training): We design a progressive training procedure for the decoder block of the U-Net generator and discriminator (See Section 3.5).
\end{itemize}
The results of this experiment are listed in Table \ref{tab:ablation}, Ablation2. Results show that the data volume can affect the performance to a large extent, and this perhaps implies why the advantages of MetroGAN are narrowed in the 256$\times$256 dataset. The geographical constraint proves to be helpful, which leads to an improvement of 4.5\%, 5.3\%, 5.0\%, and 4.1\%  with respect to the four metrics. This is because with the geographical constraint loss, we directly tell the model not to generate urban areas on the water, and the model just needs to learn the influence of water areas on surrounding regions. The progressive training procedure also brings 6.4\%, 5.0\%, 3\%, and 8.9\%  improvements in terms of four metrics respectively, indicating that the progressive growing structure can stably improve the model in all aspects.

\section{Conclusions and Future Work}
In this research, we propose the MetroGAN model, which is informed with geographical knowledge, to solve the instability problem of the previous GAN-based model for city simulation tasks, and we validate MetroGAN's ability to incorporate various urban attributes (e.g., input physical geography singly). In addition, because of cities' highly structural and hierarchical characteristics, any single metric is insufficient to evaluate generated cities. We propose a comprehensive evaluation framework for assessing whether a generated city is like a real city and how similar the generated city and the corresponding real city are, which makes up the shortcomings of the previous city extraction evaluation standards. With the new evaluation standards, we validated the superiority of our model on three baselines. Through ablation study, we show the comparison of mainstream GAN stabilizing techniques and the significant effect of geographical loss, the volume of the urban dataset, and progressive training in urban form simulation tasks. Finally, we found that with physical geographic features (terrain and water area), our model can still generate city morphology under certain natural conditions, inspiring urban geography to figure out the relationship between nature and the city. 

In future works, we plan to extend our model both spatially and temporally. From spatial perspectives, we can utilize more complex inputs into the model, such as Industrial structures and their spatial configurations, and consider urban systems with a larger spatial scale. Temporally, we plan to add recurrent neural networks into our model like \cite{ravuriSkilfulPrecipitationNowcasting2021} to 
simulate the evolution process of urban systems. Futhermore, we intend to apply interpretation methods to MetroGAN to figure out how urban attributes influence the city, and data augmentation approaches will also be added to further solve the problem of data shortage.
\begin{acks}

This research was funded by the National Natural Science Foundation of China (41830645, 41971331). Dr. Di Zhu is supported by the Faculty Set-up Funding of College of Liberal Arts, University of Minnesota (1000-10964-20042-5672018).

\end{acks}

\bibliographystyle{ACM-Reference-Format}
\balance
\bibliography{MetroGAN_rebiber}


\clearpage
\clearpage
\nobalance
\appendix

\section{Implementation Details}
\subsection{Dataset Details}
There are a few datasets for urban morphology simulation, while all of them are inaccessible, and none of them includes terrains maps. Therefore, we build and publish a new dataset for the urban morphology simulation task, containing DEM, NTL, water area, and built-up area data. These data are collected from seven data sources (See Table \ref{tab:datasource}). 
The procedures of building the dataset are as follows:

\begin{enumerate}[leftmargin=*]
    \item We calculate the total population of level 2 administrative divisions (e.g., prefecture-level cities in China) with the GADM administrative area dataset and the LandScan population density map. 
    \item We select the cities whose people are more than 10,000 and choose the center point of the cities by finding the pixel with maximum population density in the corresponding administrative district. 
    \item For every selected city, we clip an image using a rectangle window of 100km width and height (100km geographical distance corresponds to different pixels in different latitudes) around the center point. 
    \item Because the DMSP-OLS NTL data and the NPP-VIIRS NTL data have different spatial resolutions, we build two datasets for them. We resize images in the dataset of DMSP-OLS to 128$\times$128, and the images in the dataset of NPP-VIIRS to 256$\times$256 (for NPP-VIIRS). 
    \item Then, we filter the images whose amount of signal is less than 1\% (bright pixels in the built-up area account for less than 1\%). We randomly chose 200 cities as the test set.
    \item Since after filtering, only about 6000 cities are left in a single-year dataset, which is not enough for GAN's training, we collect 2000 and 2014 data to build 128$\times$128 dataset (The built-up area dataset was published in 1975,1990,2000,2014), obtaining 11773 cities. But in the 256 $\times$ 256 dataset, the NPP-VIIRS dataset was published from 2012, and only the 2014 built-up area dataset is in the period. And thus, we have to build the 256$\times$256 dataset with only 2014 data.
\end{enumerate}

\begin{figure}[htbp]
    \centering
    \begin{subfigure}[b]{\linewidth}
        \raggedright 
        \includegraphics[width=\linewidth]{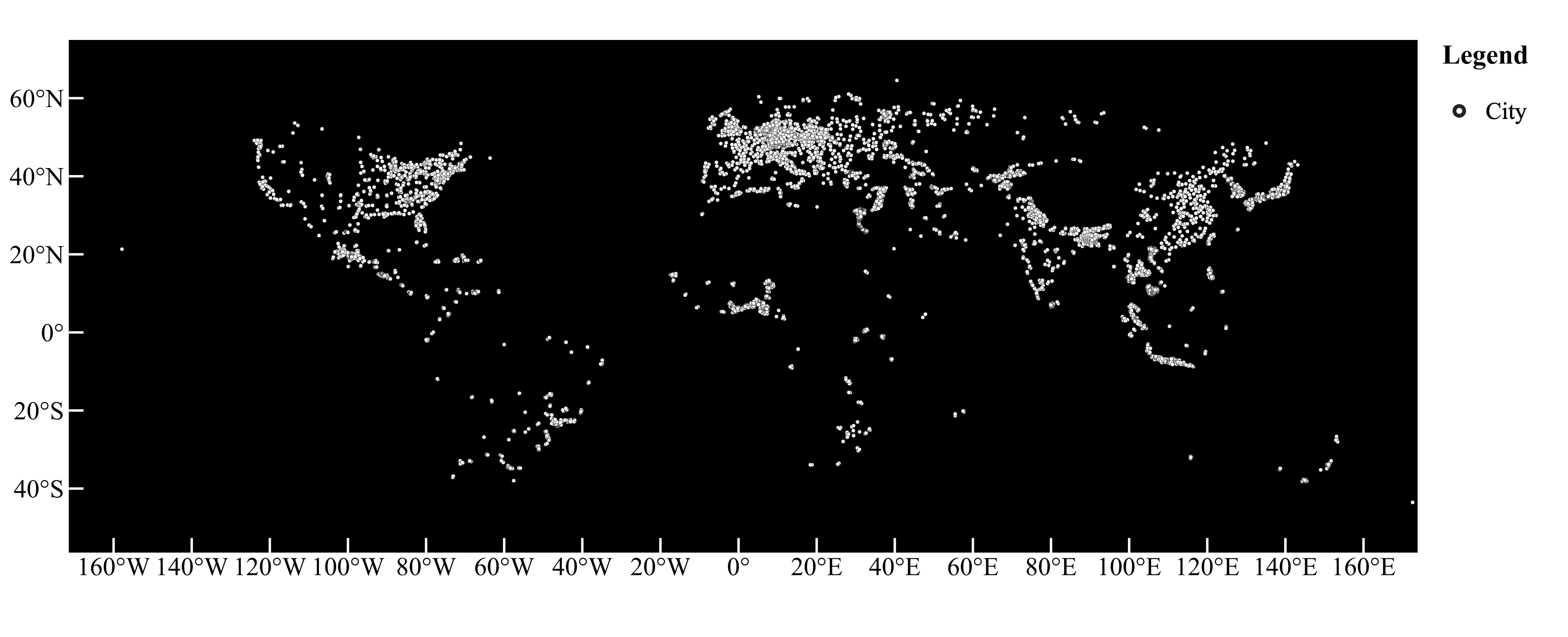}
        \caption{Distribution of global cities }
    \end{subfigure}
    \hfill
    \begin{subfigure}[b]{\linewidth}
        \raggedright 
        \includegraphics[width=1.02\linewidth]{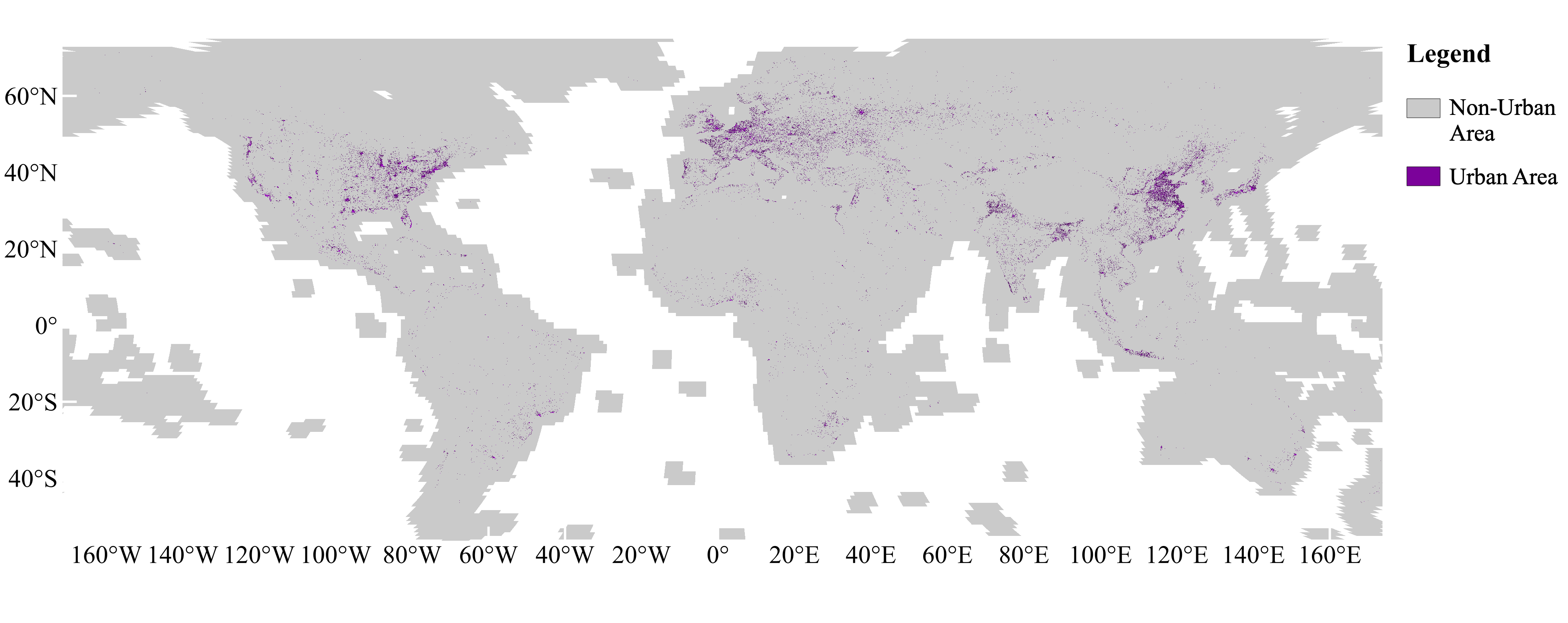}
        \caption{Distribution of built-up area }
    \end{subfigure}
    \hfill
    \begin{subfigure}[b]{\linewidth}
        \raggedright 
        \includegraphics[width=\linewidth]{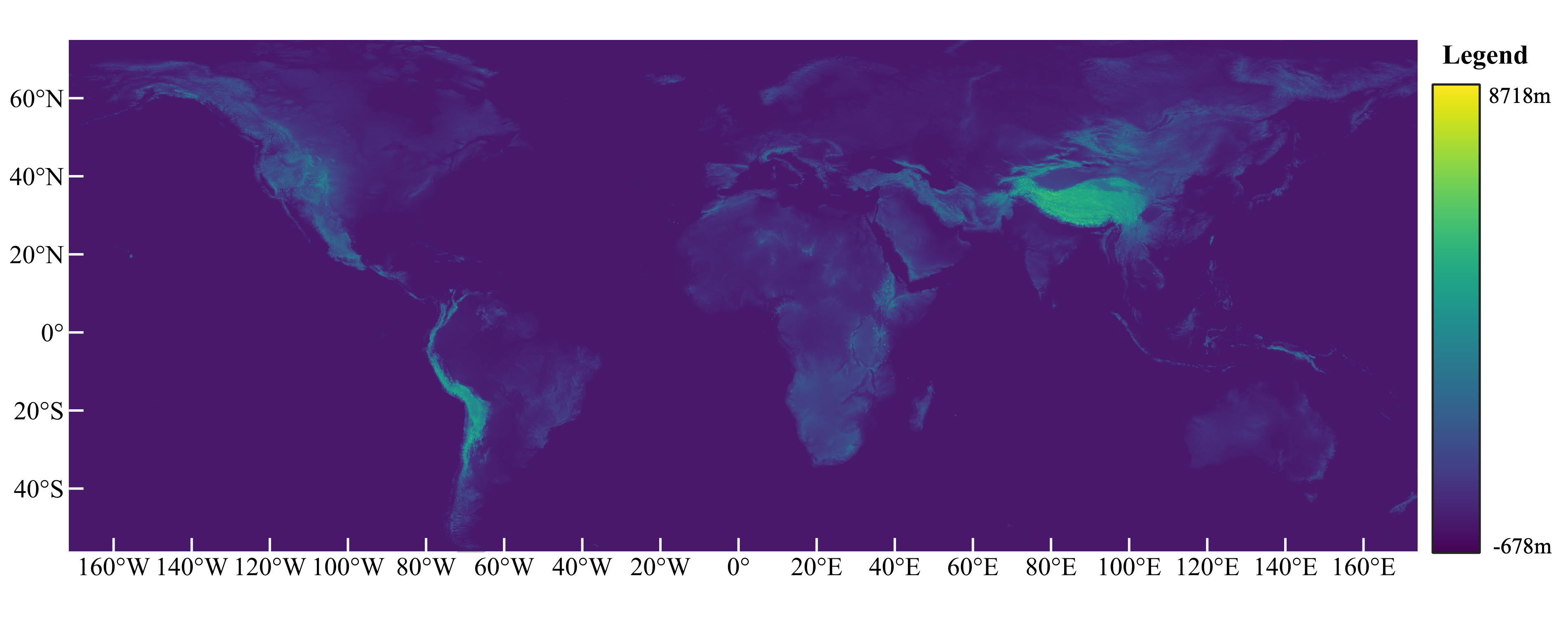} 
        \caption{Distribution of DEM}
    \end{subfigure}
    \begin{subfigure}[b]{\linewidth}
        \raggedright 
        \includegraphics[width=\linewidth]{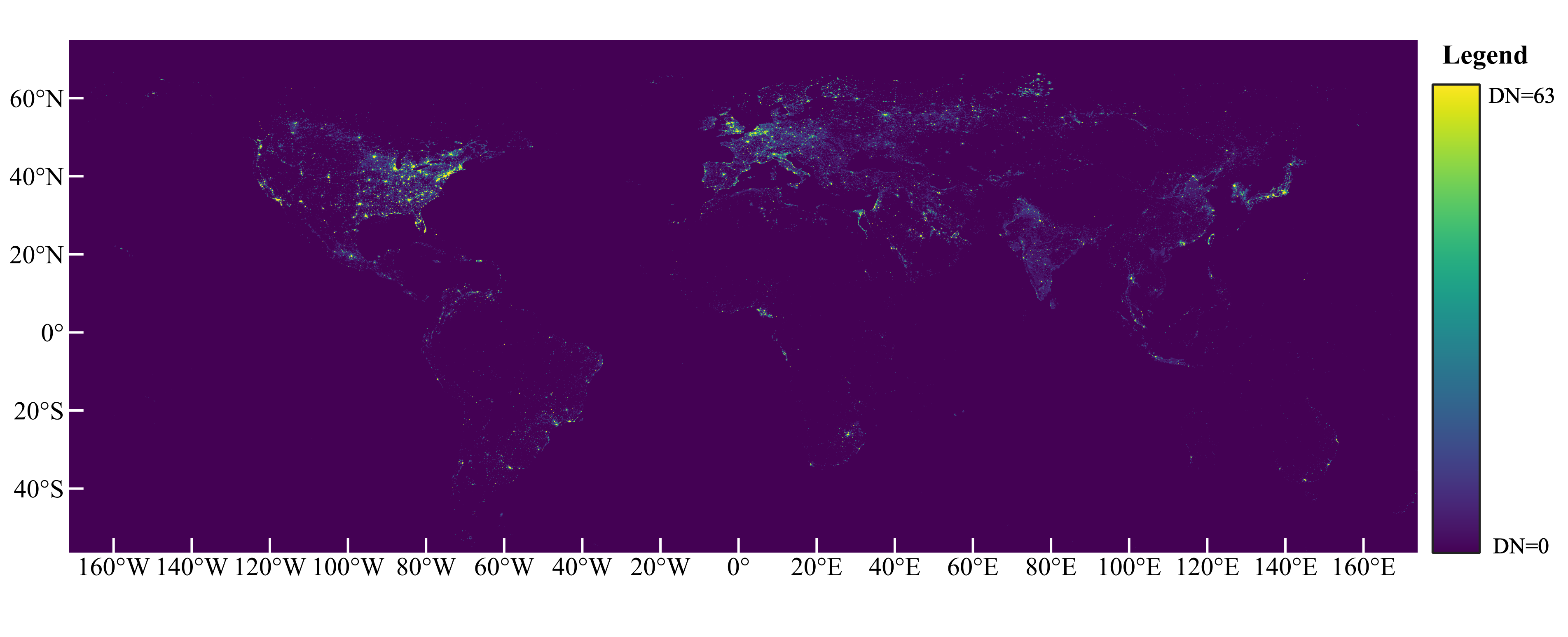}
        \caption{Distribution of nighttime light}
    \end{subfigure}
    \caption{Global distribution of the data sources}
    \label{fig:data distribution}
\end{figure}

\begin{figure}[htbp]
    \centering
    \begin{subfigure}[b]{0.8\linewidth}
        \centering
        \includegraphics[width=\linewidth]{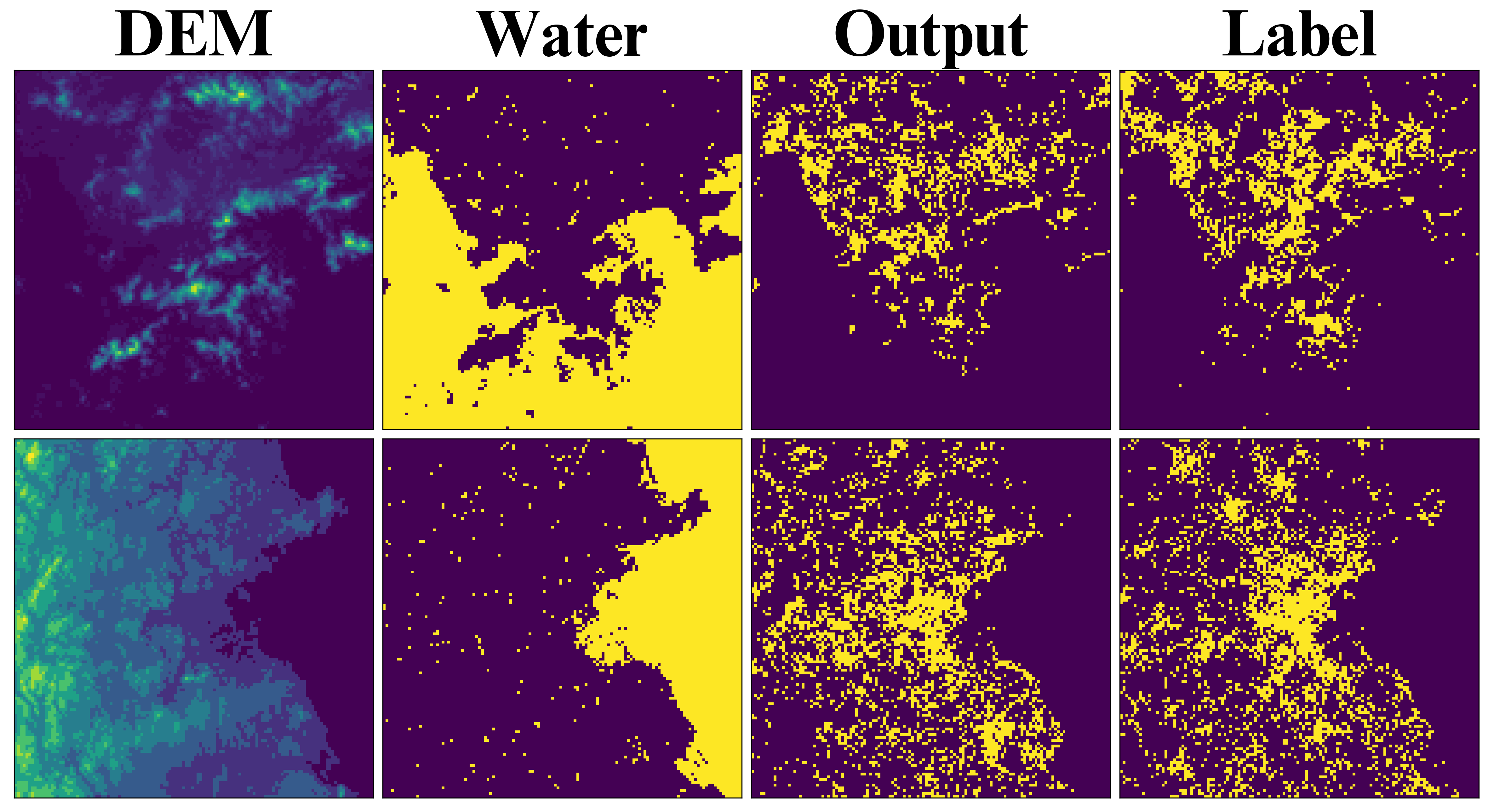}
        \subcaption{Type 1: Restricted by the sea}
    \end{subfigure}
    \hfill
    \begin{subfigure}[b]{0.8\linewidth}
        \centering
        \includegraphics[width=\linewidth]{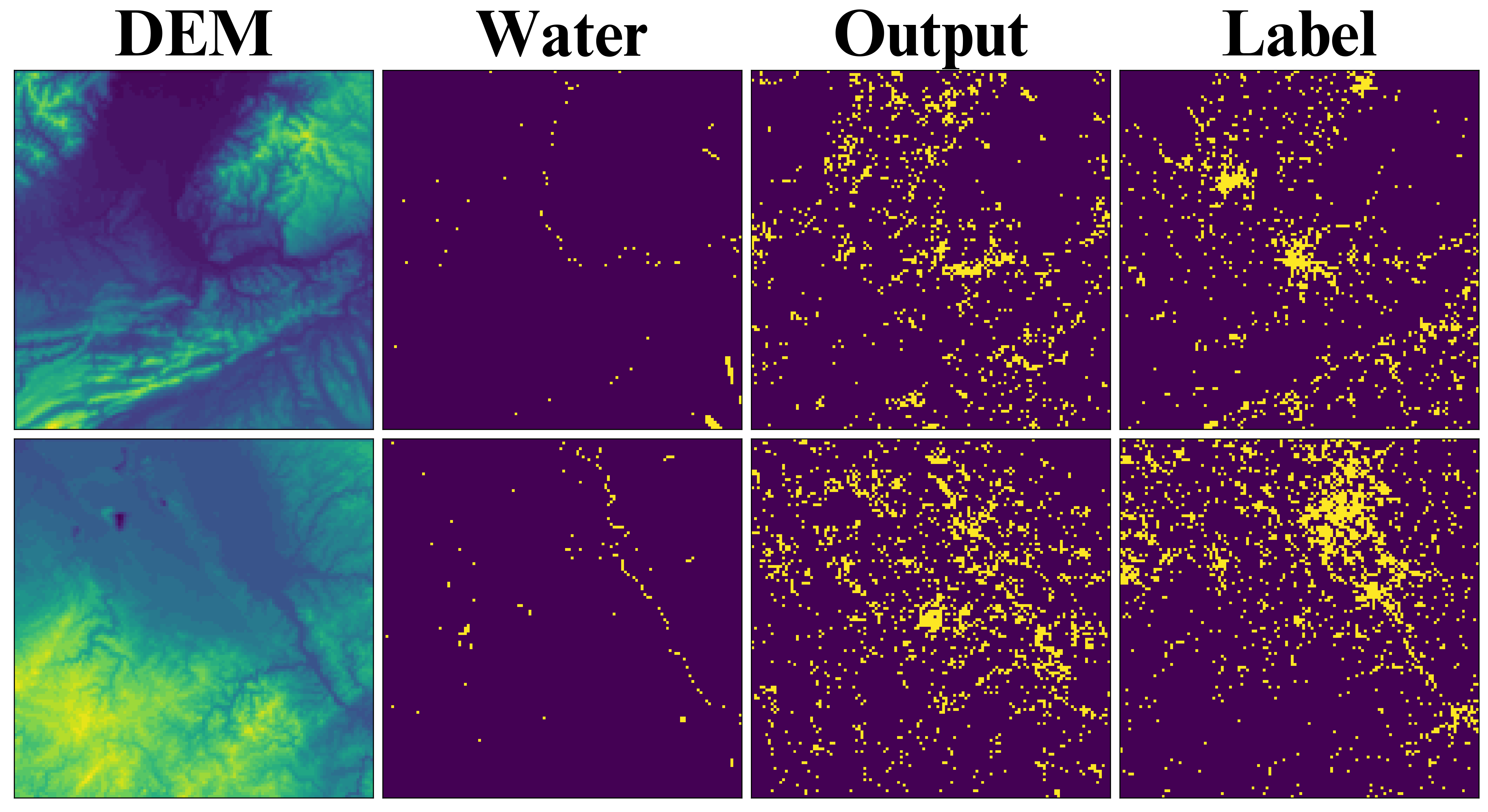}
        \caption{Type 2: Guided by the river}
    \end{subfigure}
    \hfill
    \begin{subfigure}[b]{0.8\linewidth}
        \centering
        \includegraphics[width=\linewidth]{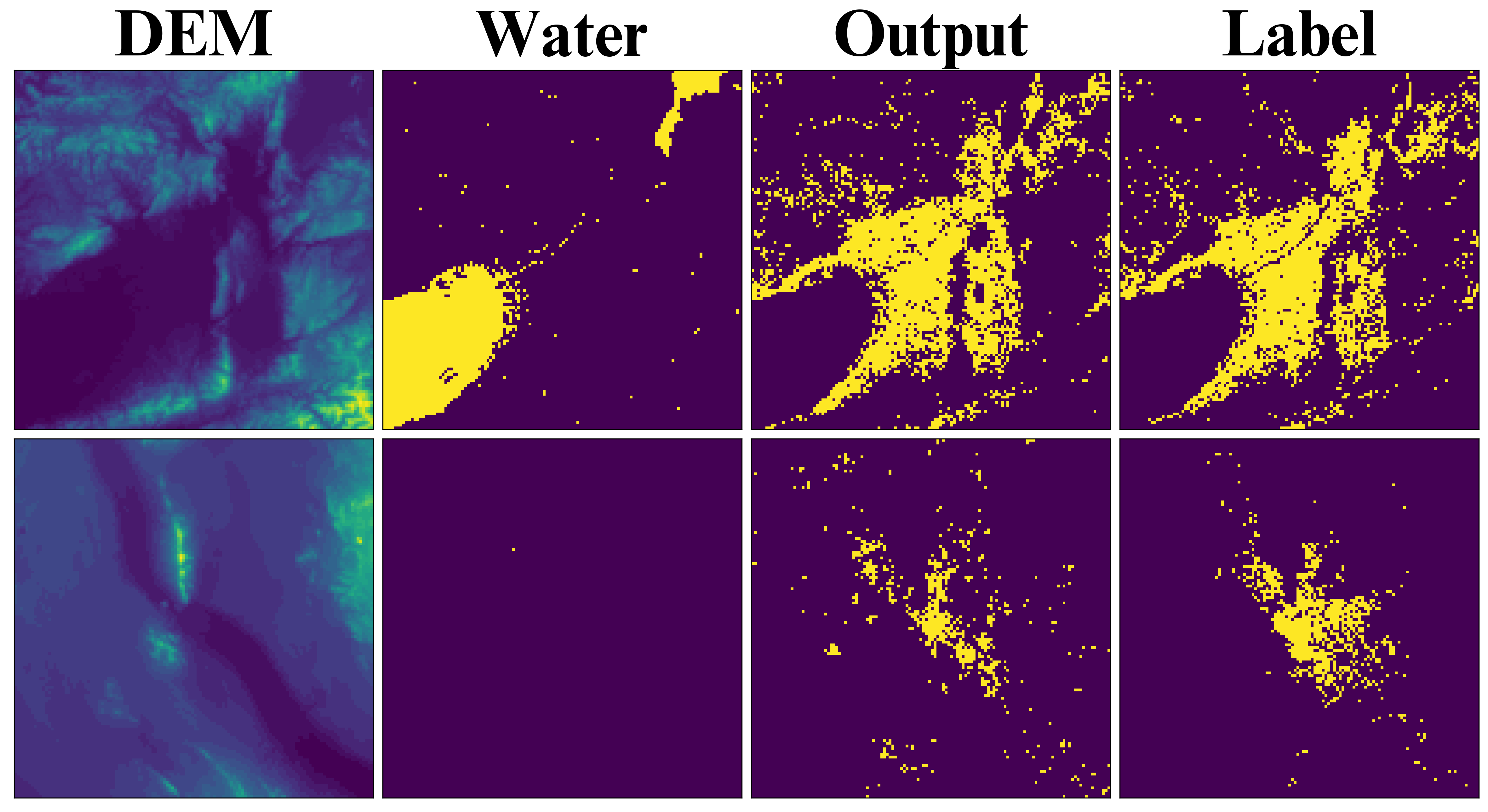} 
        \caption{Type 3: Restricted by terrains}
    \end{subfigure}
    \begin{subfigure}[b]{0.8\linewidth}
        \centering
        \includegraphics[width=\linewidth]{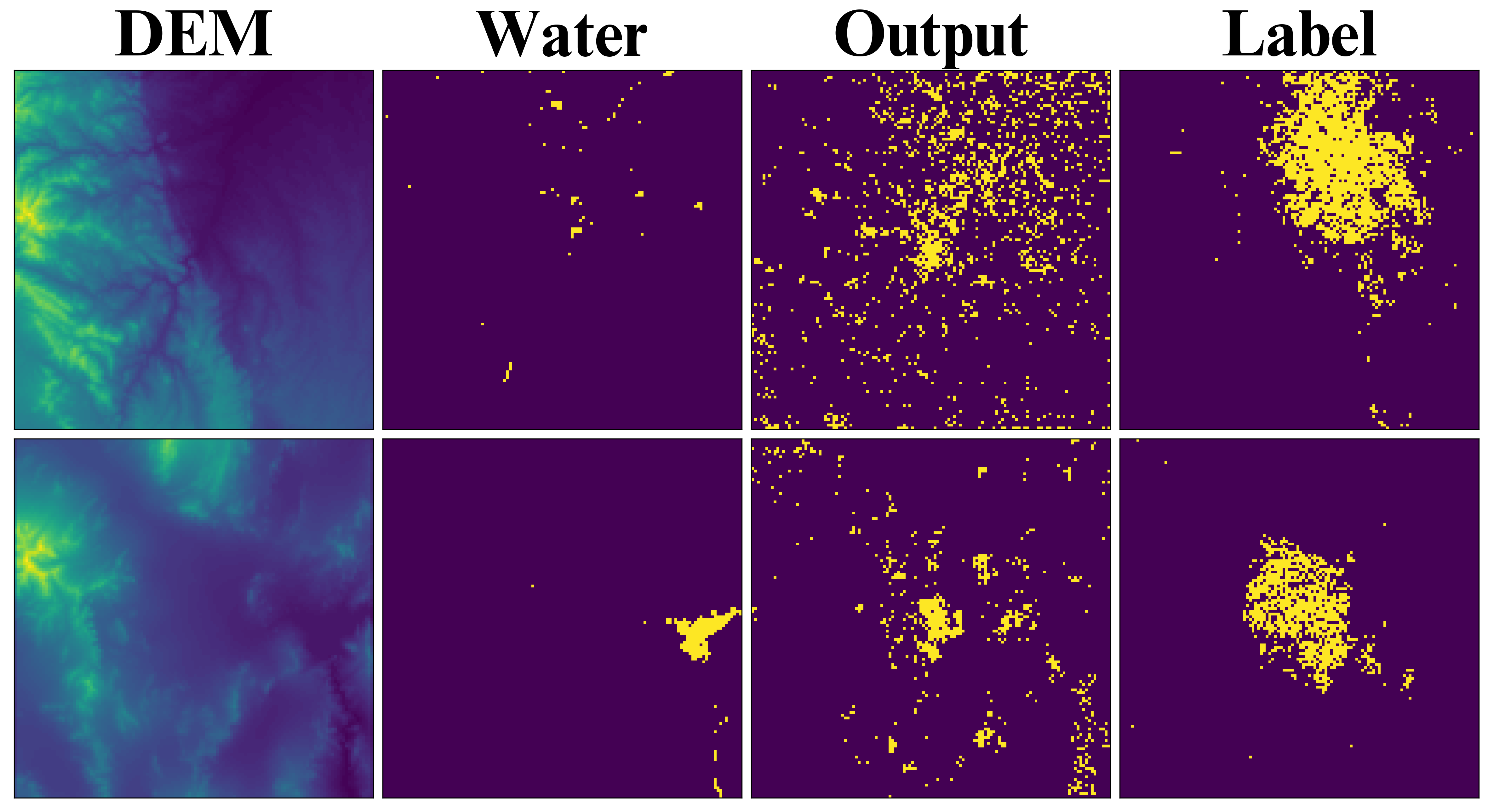}
        \caption{Type4: Few restrictions}
    \end{subfigure}
    \caption{Classification of Results of Physical Geography}
    \label{fig:nature classification}
\end{figure}
\subsection{Model Details}
\paragraph{Architecture Selection:}
In the GAN-based models training processes, the discriminator is very easy to overfit because classifying tasks of D is easier to learn than generating tasks of G. To address this problem, we set the architecture of the discriminator much simpler than the generator.

\paragraph{Progressive Growing Training Procedure:}
We add a progressive growing structure to the decoder block of the U-Net generator and discriminator. The structure of the encoder block of G is static, and the new layers of decoder block and discriminator fade in. We set the weight of the new layer $\alpha$ starts from 0 and gradually grows to 1. 

Moreover, the StyleGAN points out that only the growing G is enough, and when G is growing, the pixel-by-pixel loss is used, and other losses are not used until G grows to the highest layers. We tested this thought and found that using only the growing decoder and just L1 loss when it is growing does not decline the performance of MetroGAN. The adversarial loss and geographical loss are added after the G is grown to the highest level.

\subsection{Experimental Settings and Training Details}
We have two experimental settings. One is using NTL, DEM, and water area data as three inputs. The three inputs are all distribution maps of corresponding attributes, and the NTL data can specify the stage of the city. The other setting is preserving the physical geography inputs (DEM and water area) but without NTL data. In this setting, we need to input additional city property indicators to indicate the stage of cities. Otherwise, in theory, the model can generate the urban morphology of any historical phase at this location. Here, we use the total population of the cities and input it as a single value image. 

For hyperparameters, We tested the value of $\lambda_{L1}$ and $\lambda_{geo}$, and find when their values are 50 and 100, we can get satisfactory results. In all experiments, we set the batch size as 64, and we adopt Cosine Annealing with Warm Retarts learning rate scheduler and Adam optimizer(b1=0.5, b2=0.99). The number of iterations is 750K, and the initial learning rate is $10^{-4}$. The output of the generator is an image whose pixels' value ranges in [0,1], and we cut off the output with a threshold of 0.9.

All of our experiments are conducted on a server with two NVIDIA TITAN RTX GPUs with 24GB RAM and one Intel(R) Xeon(R) Silver 4210 CPU @ 2.20GHz on Ubuntu 18.04. The MACs of U-Net and generator in CityGAN and MetroGAN are 13.34G. The MACs of discriminator in CityGAN and MetroGAN are 1.28G. The MACs of these models are comparable, and MetroGAN shows its superiority with better performances. The source code and the dataset are available at Github (https://github.com/zwy-Giser/MetroGAN).
\section{ADDITIONAL ANALYSIS OF RESULTS}
\subsection{Analysis of results with physical geography inputs}
In the experiments with physical constraints input, we find that in special conditions, the urban morphology can be simulated vividly by our model. Specifically, we qualitatively summarize the conditions into four types. 
\begin{itemize}[leftmargin=*]
\item Restricted by the sea. Most cities by the sea can be well modeled. This is because one or two sides of the urban morphology have been restricted by the coast. Besides, the urban center of coastal cities is always along the coast, and the other side of the coast may be restricted by terrains, such as mountains.
\item Guided by the river. In some cases that contain rivers, the spines of the urban morphology follow the flow of the river. 
\item Restricted by terrains. When the restrictions of terrains are strong, such as in a basin, the model can generate more accurate urban forms because available space is limited.
\item Few Restrictions. When there are few restrictions on water and terrains, such as on a plain, the urban growth is likely driven by other factors. The most significant difference in these cases is that the real cities reflect strong agglomeration effects, but simulated cities do not. This is because our model does not consider the spatial interaction process.
\end{itemize}
\subsection{Outliers of Macroscopic Level Validation}
This section shows some outliers of fractal dimension validation. The outliers are picked with the standard that the difference between FD of the real city and FD of the generated city is larger than 0.2. And we find that these outliers all have a small number of bright pixels(less than 15\% pixels). The small amount of urban area means the calculation of these cities' fractal dimensions has low sensitivity. Only a few wrongly classified bright pixels can affect the fractal dimension to a large extent.
\begin{figure}[h]
    \centering
    \includegraphics[width=\linewidth]{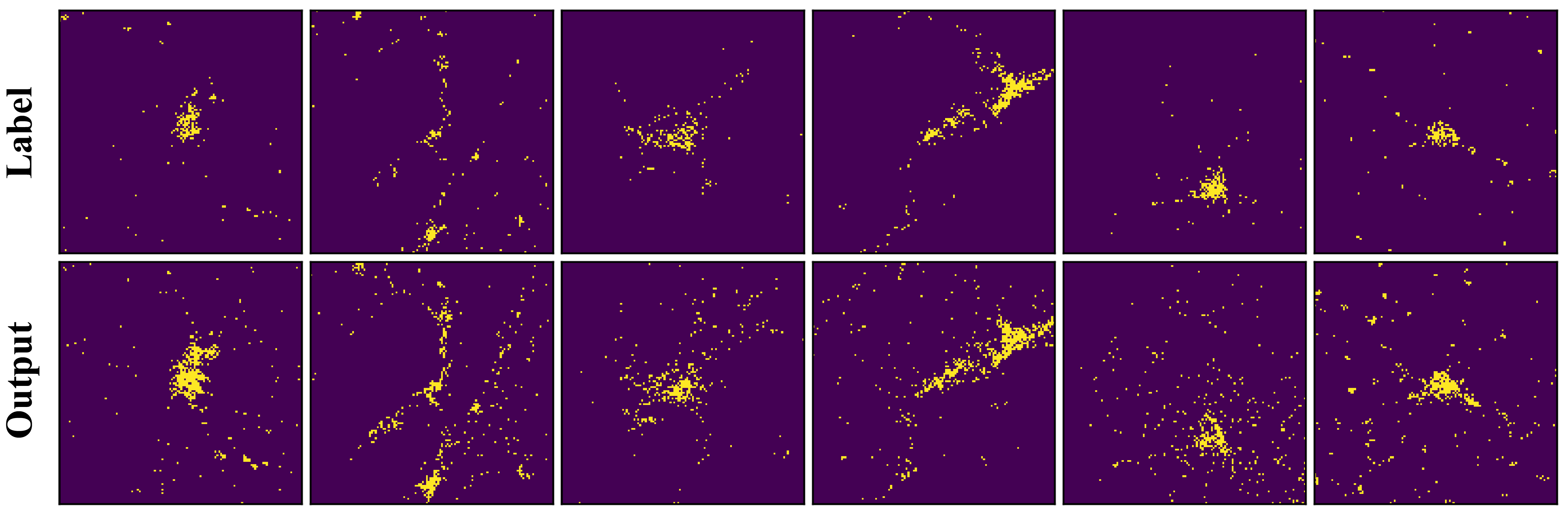}
    \caption{Outliers of fractal dimension validation}
    \label{fig:outliers}
\end{figure}
\end{document}